\documentclass[usegraphicx,usenatbib,letterpaper]{emulateapj}

\usepackage{ulem}
\usepackage{color}
\usepackage{graphics,graphicx}
\usepackage{times} 
\usepackage{amssymb}
\usepackage{amsmath}
\usepackage{natbib}
\usepackage{url}
\usepackage{multirow}
\usepackage{ctable}
\usepackage{threeparttable}
\newif\ifAMStwofonts
\AMStwofontstrue

\def\xmm{{\it XMM-Newton}}

\def\suzaku{{\it Suzaku}}

\def\chandra{{\it Chandra}}

\def\epicpn{{\it EPIC}{\rm-pn}}
\def\epicmos1{{\it EPIC}{\rm-MOS1~\/}}
\def\epicmos2{{\it EPIC}{\rm-MOS2 ~\/}}
\def\epicmos{{\it EPIC}{\rm-MOS}}

\def\nustar{{\it NuSTAR}}
\def\athena{{\it ATHENA+}}
\def\astroh{{\it Astro--H}}

\def\deg{$^{\circ}$}

\def\H0{{\rm ~km~s^{-1}~Mpc^{-1}}}

\def\kev{\hbox{\rm keV}}

\def\atpcm{{\rm atom~cm$^{-2}$}}

\def\ergcmps{\hbox{\rm erg~cm~s$^{-1}$}}

\def\msun{\hbox{$\rm M_{\odot}$}}

\def\chisq{{$\chi^{2}$}}

\def\xspnorm{{\rm ct~cm$^{-2}$~s$^{-1}$~keV$^{-1}$}}
\def\xspec{\hbox{\small XSPEC}}

\def\heasoft{\hbox{\rm{\small HEASOFT}}}
\def\nustardas{\rm {\small NUSTARDAS}}
\def\xselect{\hbox{\rm{\small XSELECT}}}
\def\xmmselect{\hbox{\rm{\small XMMSELECT}}}
\def\ftool{\hbox{\rm{\small FTOOL}}}

\def\specgroup{\rm {\small SPECGROUP}}

\def\addascaspec{\hbox{\rm{\small ADDASCASPEC~\/}}}
\def\sas{\hbox{\rm{\small SAS~\/}}}
\def\epchain{\hbox{\rm{\small EPCHAIN}}}
\def\emchain{\hbox{\rm{\small EMCHAIN}}}

\def\rmfgen{\hbox{\rm{\small RMFGEN}}}
\def\arfgen{\hbox{\rm{\small ARFGEN}}}
\def\epiclccorr{\hbox{\rm{\small EPICLCCORR}}}

\def\addascaspec{\hbox{\rm{\small ADDASCASPEC}}}

\def\xstar{\hbox{\rm{\small XSTAR}}}
\def\grid25{\hbox{\rm{\small GRID25}}}

\def\reflionx{\rm{\small REFLIONX}}
\def\xillver{\rm{\small XILLVER}}

\def\relconv{\rm{\small RELCONV}}
\def\relline{\rm{\small RELLINE}}

\def\pexrav{\rm{\small PEXRAV}}

\def\pcfabs{\rm{\small PCFABS}}

\def\ka{K\,$\alpha$}
\def\kb{K\,$\beta$}

\def\etal{et al.}
\def\eg{{\it e.g.~\/}}

\def\ie{{\it i.e.~\/}}

\def\la{\mathrel{\hbox{\rlap{\hbox{\lower4pt\hbox{$\sim$}}}{\raise2pt\hbox{$<$}}}}}
\def\ga{\mathrel{\hbox{\rlap{\hbox{\lower4pt\hbox{$\sim$}}}{\raise2pt\hbox{$>$}}}}}

\def\d25{D$_{25}$}
\def\nh{{$N_{\rm H}$}}

\def\.25{0.25 keV\thinspace}

\def\rg{$R_{\rm G}$}

\def\ngc{\rm NGC\,1365}

\shorttitle{NGC\,1365: Spectral Variability and Black Hole Spin}
\shortauthors{D.~J. Walton et al.}

\begin{document}

\title{\nustar\ and \xmm\ Observations of NGC\,1365: Extreme Absorption Variability and a Constant Inner Accretion Disk}

\author{D. J. Walton\altaffilmark{1},
G. Risaliti\altaffilmark{2,3},
F. A. Harrison\altaffilmark{1},
A. C. Fabian\altaffilmark{4},
J. M. Miller\altaffilmark{5},
P. Arevalo\altaffilmark{6},
D. R. Ballantyne\altaffilmark{7},
S. E. Boggs\altaffilmark{8},
L. W. Brenneman\altaffilmark{3}, 
F. E. Christensen\altaffilmark{9},
W. W. Craig\altaffilmark{8},
M. Elvis\altaffilmark{3},
F. Fuerst\altaffilmark{1},
P. Gandhi\altaffilmark{10},
B. W. Grefenstette\altaffilmark{1},
C. J. Hailey\altaffilmark{11},
E. Kara\altaffilmark{4},
B. Luo\altaffilmark{12,13},
K. K. Madsen\altaffilmark{1},
A. Marinucci\altaffilmark{14},
G. Matt\altaffilmark{14},
M. L. Parker\altaffilmark{4},
C. S. Reynolds\altaffilmark{15,16},
E. Rivers\altaffilmark{1},
R. R. Ross\altaffilmark{17},
D. Stern\altaffilmark{18},
W. W. Zhang\altaffilmark{19}
}
\affil{
$^{1}$Cahill Center for Astronomy and Astrophysics, California Institute of Technology, Pasadena, CA 91125, USA \\
$^{2}$INAF -- Osservatoria Astrofisico di Arcetri, Largo Enrico Fermi 5, 50125 Firenze, Italy \\
$^{3}$Harvard-Smithsonian Center for Astrophysics, 60 Garden Street, Cambridge, MA 02138, USA \\
$^{4}$Institute of Astronomy, University of Cambridge, Madingley
Road, Cambridge CB3 0HA, UK \\
$^{5}$Department of Astronomy, University of Michigan, 500
Church Street, Ann Arbor, MI 48109-1042, USA \\
$^{6}$Pontificia Universidad Cat\'olica de Chile, Instituto de Astrf\'isica, Casilla 306, Santiago 22, Chile \\
$^{7}$Center for Relativistic Astrophysics, School of Physics, Georgia Institute of Technology, Atlanta, GA 30332, USA \\
$^{8}$Space Sciences Laboratory, University of California, Berkeley, CA 94720, USA \\
$^{9}$DTU Space, National Space Institute, Technical University of Denmark, Elektrovej 327, DK-2800 Lyngby, Denmark \\
$^{10}$Department of Physics, Durham University, South Road, Durham DH1 3LE, UK \\
$^{11}$Columbia Astrophysics Laboratory, Columbia University, New York, NY 10027, USA \\
$^{12}$Department of Astronomy and Astrophysics, 525 Davey Lab, The Pennsylvania State University, University Park, PA 16802, USA \\
$^{13}$Institute for Gravitation and the Cosmos, The Pennsylvania State University, University Park, PA 16802, USA \\
$^{14}$Dipartimento di Matematica e Fisica, Universita degli Studi Roma Tre, via della Vasca Navale 84, 00146 Roma, Italy \\
$^{15}$Department of Astronomy, University of Maryland, College Park, MD 20742, USA \\
$^{16}$Joint Space-Science Institute (JSI), University of Maryland, College Park, MD 20742, USA \\
$^{17}$Physics Department, College of the Holy Cross, Worcester, MA 01610, USA \\
$^{18}$Jet Propulsion Laboratory, California Institute of Technology, Pasadena, CA 91109, USA \\
$^{19}$NASA Goddard Space Flight Center, Greenbelt, MD 20771, USA \\
}

\begin{abstract}
We present a spectral analysis of four coordinated \nustar+\xmm\ observations
of the Seyfert galaxy \ngc. These exhibit an extreme level of spectral variability,
which is primarily due to variable line-of-sight absorption, revealing relatively
unobscured states in this source for the first time. Despite the diverse range of
absorption states, each of the observations displays the same characteristic
signatures of relativistic reflection from the inner accretion disk. Through
time-resolved spectroscopy we find that the strength of the relativistic iron line
and the Compton reflection hump relative to the intrinsic continuum are well
correlated, as expected if they are two aspects of the same broadband reflection
spectrum. We apply self-consistent disk reflection models to these time-resolved
spectra in order to constrain the inner disk parameters, allowing for variable,
partially covering absorption to account for the vastly different absorption states
observed. Each of the four observations is treated independently to test the
consistency of the results obtained for the black hole spin and the disk inclination,
which should not vary on observable timescales. We find both the spin and the
inclination determined from the reflection spectrum to be consistent, confirming
\ngc\ hosts a rapidly rotating black hole; in all cases the dimensionless spin
parameter is constrained to be $a^* > 0.97$ (at 90\% statistical confidence or
better).
\end{abstract}


\section{Introduction}

Black hole spin is a quantity of significant importance for addressing a variety
of astrophysical topics, including the growth of the supermassive black holes
powering active galactic nuclei (AGN; \eg \citealt{Dubois13}), the formation of
black hole binaries in supernova explosions (\citealt{Miller11}), and potentially
the launch of powerful relativistic jets (\citealt{BZ77}), although the exact role
spin plays here is still controversial (\citealt{Steiner13, King13jet, Russell13}).
For active galaxies, the best method available for measuring black hole spin is
to measure the relativistic distortions of fluorescent line emission from the
inner accretion disk, excited through irradiation by hard X-rays. These
relativistic effects broaden and skew intrinsically narrow emission lines into a
characteristic `diskline' profile (\citealt{Fabian89, kdblur}) which depends on
black hole spin.

The most prominent features produced by such irradiation are typically the iron
\ka\ emission line at $\sim$6.4--7.0\,\kev\ (depending of ionization state) and
a broad peak in the reflected continuum at $\sim$30\,\kev\ referred to as the
Compton hump (\citealt{George91}). Roughly $\sim$40\% of X-ray bright AGN
display evidence for broadened iron \ka\ emission (\citealt{Nandra07,
dLCPerez10}), and the majority of AGN also show a `hard' excess above
$\sim$10\,\kev\ (\eg \citealt{Nandra94, Perola02, Dadina07, Rivers13}),
consistent with Compton reflection from the accretion disk (\eg
\citealt{Walton10Hex, Nardini11}). Spin estimates for a growing sample
($\sim$20--30 sources) of local AGN have recently been obtained through study
of these features, \eg \cite{kerrconv, Miniutti09sw2127, Zoghbi10, Brenneman11,
Nardini12, Gallo13, Walton13spin}; see also \cite{Miller07rev, Reynolds13rev,
Brenneman13book} for recent reviews. Current results indicate the majority of
AGN may host rapidly rotating black holes, although the sample is still fairly
small, and not yet well defined in a statistical sense.

However, the identification of these spectral features with relativistic disk
reflection is not without controversy. In particular, scenarios fully dominated by
absorption and reprocessing from material relatively distant to the AGN have
frequently been proposed as alternative interpretations to relativistic disk 
reflection (\eg \citealt{Miller08, LMiller09, Sim10}). If these absorbing and
scattering structures are allowed sufficiently complex geometries, such models
are also able to reproduce the observed X-ray spectra. Many AGN do indeed
display evidence for partially ionised absorption in X-rays (\eg
\citealt{Blustin05}), as well as evidence of reflection from distant, dense
material in the form of a  narrow iron \ka\ emission line (\eg \citealt{Bianchi09}),
and a number also display evidence for variable absorption (\eg
\citealt{Risaliti02}). The contributions of this absorption can be difficult to fully
disentangle from relativistic disk reflection in some cases, particularly without
access to sensitive hard X-ray ($>$10\,\kev) coverage.

\begin{table}
  \caption{Basic Observational Details for the Coordinated \nustar+\xmm\
observations of \ngc.}
\begin{center}
\begin{tabular}{c c c c}
\hline
\hline
\\[-0.2cm]
Observation & Date & \multicolumn{2}{c}{Total Good Exposures (ks)} \\
& & \xmm\tmark[a] & \nustar\tmark[b] \\
\\[-0.3cm]
\hline
\hline
\\[-0.15cm]
1 & July 2012 & 110/130 & 77 \\
\\[-0.25cm]
2 & Dec 2012 & 93/120 & 66 \\
\\[-0.25cm]
3 & Jan 2013 & 90/118 & 74 \\
\\[-0.25cm]
4 & Feb 2013 & 103/119 & 70 \\
\\[-0.25cm]
\hline
\hline
\\[-0.25cm]
\end{tabular}
\end{center}
$^{a}$ \xmm\ exposures are quoted for the \epicpn/each of the \epicmos\
detectors. \\
$^{b}$ \nustar\ exposures are quoted for each of the focal plane modules. \\
\label{tab_obs}
\end{table}

\begin{figure*}
\epsscale{1.16}
\hspace*{-0.6cm}
\vspace*{0.5cm}
\plotone{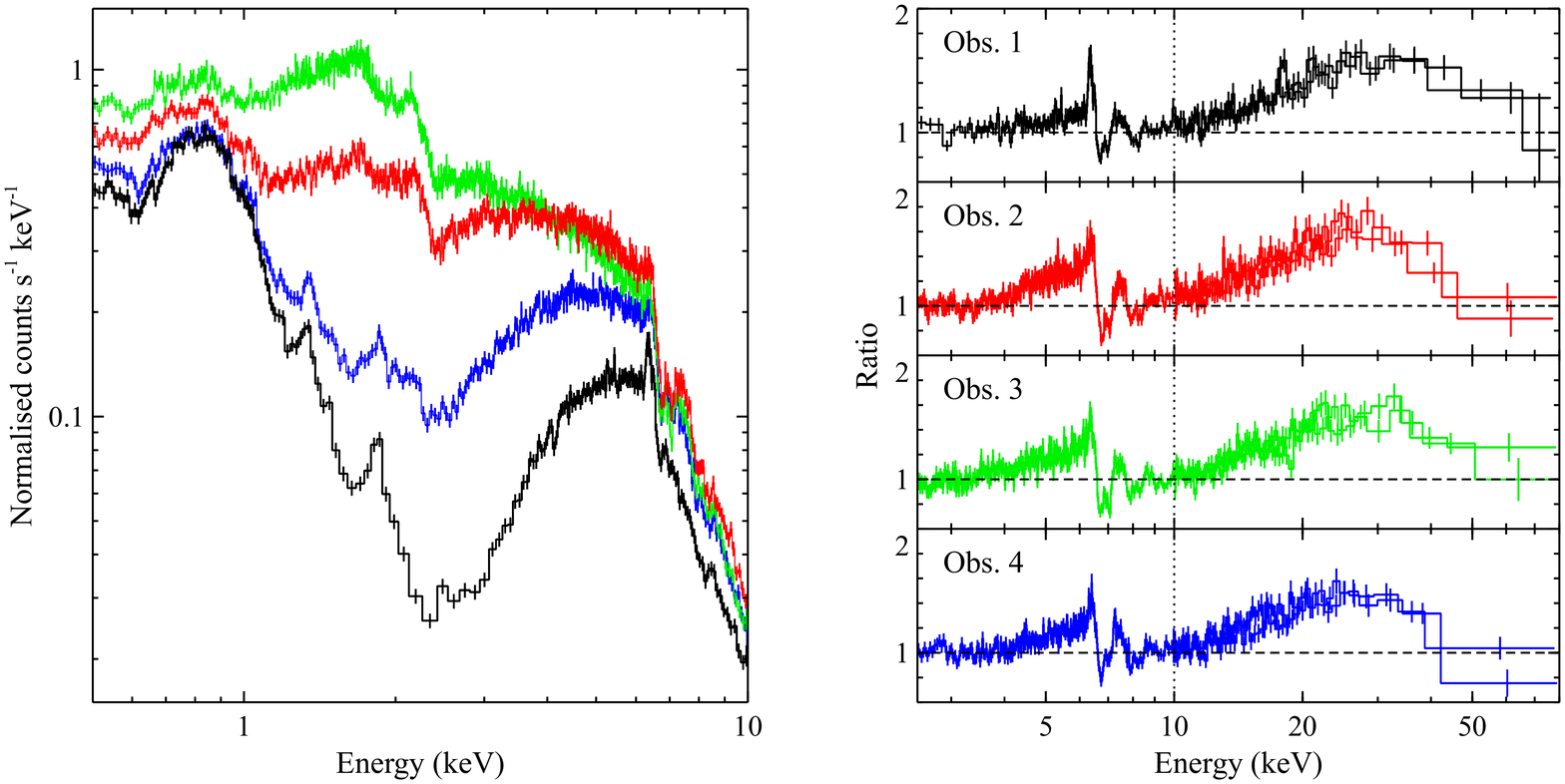}
\hspace*{-0.6cm}
\plotone{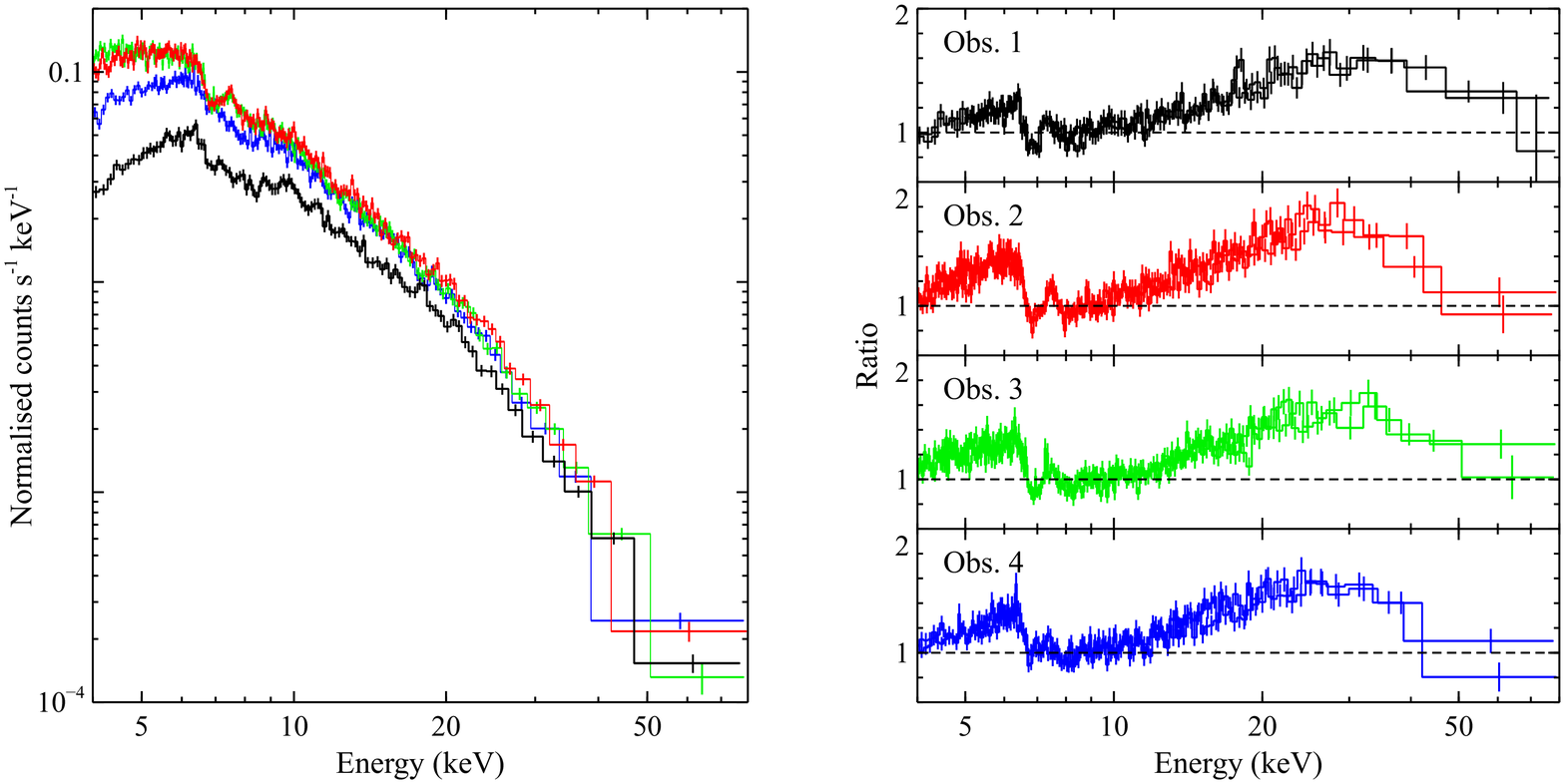}
\vspace*{0.5cm}
\caption{
\textit{Top-left panel:} time-averaged \xmm\ \epicpn\ spectra from each of the
four coordinated \nustar+\xmm\ observations of \ngc, demonstrating the
extreme spectral variability displayed. Observations 1, 2, 3 and 4 are shown in
black, red, green and blue respectively. \textit{Top-right panel:} residuals to a
simple $\Gamma = 1.75$ powerlaw continuum, modified by partially covering
neutral absorption, and applied to the 2.5--4, 7--10 and 50--80\,\kev\ energy
ranges. For clarity, we show the \xmm\ \epicpn\ data below 10\,\kev, and the
\nustar\  FPMA/FPMB data above 10\,\kev. The same hallmarks of reflection
from the inner accretion disk, \ie a relativistically broadened iron line at
$\sim$6\,\kev\ and a strong Compton hump at $\sim$30\,\kev\ are seen in
each of the four observations, despite the extreme variation in the line-of sight
absorbing column. \textit{Bottom panels:} as for the top panels, but now
displaying only the \nustar\ data, further highlighting the reduced variability at
high energies compared to that seen at $\sim$2\,\kev, and the detection of the
broad iron line in these data. The narrow component of the iron emission is
less visually prominent in the \nustar\ data owing to the coarser spectral
resolution in the iron \ka\ bandpass in comparison to \xmm. In the left panel,
only the data from FPMA is shown for clarity. The data in all panels have been
rebinned for visual purposes.
}
\label{fig_ratio}
\end{figure*}

Since the launch of the Nuclear Spectroscopic Telescope Array (\nustar;
\citealt{NUSTAR}), \ngc\ has become central to the debate over the
contribution of emission from the inner disk in AGN. \ngc\ ($z = 0.0055, D
\sim 20$\,Mpc) is a well studied Seyfert 1.9 galaxy, hosting a $\sim$2
$\times 10^{6}$\,\msun\ black hole (\citealt{Schulz99, Kaspi05}) which
displays evidence for a relativistically broadened iron line (\citealt{Risaliti09a,
Walton10Hex, Brenneman13}) indicative of a rapidly rotating black hole.
However, it is also known to display complex and variable absorption
(\citealt{Risaliti05b, Risaliti05a, Risaliti09a, Maiolino10}). The unprecedented
high energy data quality and the continuous $\sim$3--80\,\kev\ bandpass
provided by \nustar\ is ideal for the study of X-ray reflection (\eg
\citealt{Miller13grs, Tomsick14, Marinucci14}; Parker \etal\ 2014c,
\textit{submitted}). Early in the mission, \nustar\ observed \ngc, coordinated
with \xmm\ (\citealt{XMM}) for soft X-ray coverage, detecting both a
relativistic iron line and a strong hard excess. The hard X-ray data from
\nustar\ displayed excellent consistency with the prediction of the disk
reflection interpretation from the \xmm\ data, and revealed that a
Compton-thick absorber would be required to reproduce the high energy 
data without invoking disk reflection. The presence of such material was
found to be inconsistent with the levels of reprocessing observed, either from
neutral or partially ionised material (\citealt{Risaliti13nat}), providing a strong
confirmation of the contribution from relativistic disk reflection. However,
based on a set of simulated spectra, \cite{LMiller13} subsequently challenged
this conclusion, suggesting that distant absorption/reprocessing could yet
explain the spectra observed from \ngc, and claiming that AGN spins cannot
be measured at all.

Here we present results from the full series of four coordinated observations
of \ngc\ performed by \nustar\ and \xmm, the first of which was initially
presented in \cite{Risaliti13nat}. These observations probe an unprecedented
range of absorption states with high signal-to-noise data, and also reveal a
persistent contribution from the inner accretion disk. We use these observations
to disentangle the relative contribution of these processes, and determine the
inner disk parameters. This work is structured as follows: the data reduction
procedure is outlined in section \ref{sec_red}, our analysis is presented in
section \ref{sec_spec}, and the results obtained discussed in section
\ref{sec_dis}. Finally, we summarize our conclusions in section~\ref{sec_conc}.

\section{Data Reduction}
\label{sec_red}

\nustar\ and \xmm\ have performed four coordinated observations of \ngc\ to
date, in July and December 2012, and January and February  2013, hereafter
observations 1--4 respectively. Basic observation details are given in Table
\ref{tab_obs}. Here we outline our data reduction procedure for these observations.

\subsection{NuSTAR}

The \nustar\ data were reduced using the standard pipeline, part of the \nustar\
Data Analysis Software v0.11.1 (\nustardas; now part of the standard \heasoft\
distribution), and instrumental responses from \nustar\ caldb v20130509 are
used throughout this work. The unfiltered event files were cleaned with the
standard depth correction, which significantly reduces the internal background at
high  energies, and South Atlantic Anomaly passages were excluded from our
analysis. Source products were obtained from circular regions (radius
$\sim$100--120$''$), and background was estimated from a blank area of the
same detector free of contaminating point sources. Spectra and lightcurves were
extracted from the cleaned event files using \xselect\ for both focal plane
modules (FPMA and FPMB). Finally, the spectra were grouped such that each
spectral bin contains at least 50 counts. The data from FPMA and FPMB are
analyzed jointly in this work, but are not combined together.

\subsection{XMM-Newton}

Data reduction was carried out with the \xmm\ Science Analysis System (\sas
v13.0.0) largely according to the standard prescription provided in the online
guide\footnote{http://xmm.esac.esa.int/}. The observation data files were
processed using \epchain\ and \emchain\ to produce calibrated event lists for
the \epicpn\ (\citealt{XMM_PN}) and \epicmos\ (\citealt{XMM_MOS}) detectors
respectively. Source products were extracted from a circular region of
$\sim$40$''$ in radius, and background was estimated from an area of the
same CCD free of contaminating point sources. Lightcurves and spectra were
generated with \xmmselect, selecting only single and double events (single to
quadruple events) for \epicpn\ (\epicmos), excluding periods of high
background flares (occuring predominantly at the end of the observations). The
redistribution matrices and auxiliary response files were generated with
\rmfgen\ and \arfgen, while lightcurves were corrected for the background
count rate using \epiclccorr. After performing the data reduction separately for
each of the MOS CCDs and confirming their consistency, the spectra were
combined using the \ftool\ \addascaspec. Finally, spectra were re-binned using
the SAS task \specgroup\ to have a minimum signal-to-noise (S/N) of 5 in each
energy bin.

\section{Spectral Analysis}
\label{sec_spec}

Throughout this work, spectral analysis is performed with \xspec\ v12.8.0
(\citealt{XSPEC}), and unless stated otherwise parameter uncertainties are
quoted at the 90\% level of confidence for a single parameter of interest (\ie
$\Delta$\chisq\ = 2.71). The spectral agreement between \xmm\ and \nustar\
is generally good across their common energy range ($\sim$3--10\,\kev; \eg
\citealt{Walton13culx, Walton14hoIX}; Madsen et al., \textit{in preparation}), but
we treat residual uncertainties in the flux calibration between the various
detectors utilized by allowing variable constants to float between them, fixing
the constant for the \nustar\ FPMA data to unity.

\begin{figure*}
\hspace*{-0.35cm}
\epsscale{1.145}
\plotone{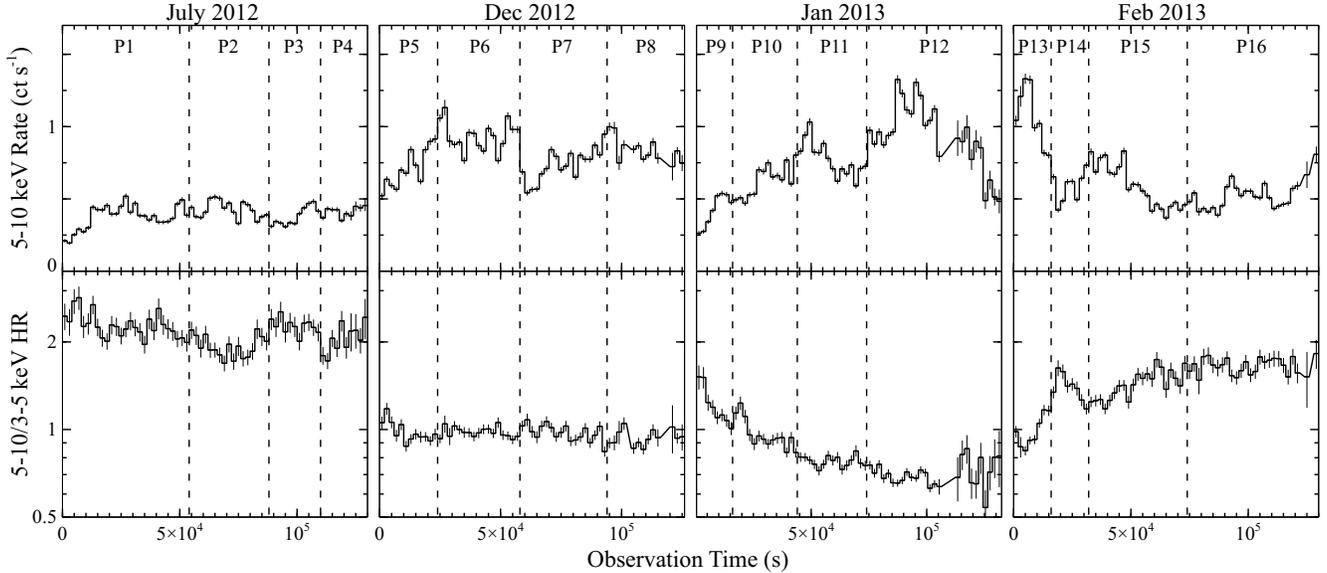}
\caption{
\textit{Top panels:} The 5--10\,\kev\ \xmm\ \epicpn\ lightcurves observed
from \ngc\ for each of the four observations. \textit{Bottom panels:} the
5--10/3--5\,\kev\ hardness ratio lightcurves. The vertical dashed lines indicate
the 16 periods selected for our time-resolved analysis (4 periods are defined
within each observation), based on either changes in the 5--10\,\kev\ lightcurve,
and/or the 5--10/3--5\,\kev\ hardness ratio.
}
\hspace*{0.5cm}
\label{fig_lc}
\end{figure*}

\subsection{Extreme Spectral Variability}

It is clear from a simple visual comparison of the average spectra obtained from
the four observations, shown in Figure \ref{fig_ratio} (\textit{left panels}), that
our coordinated program caught a truly extreme level of spectral variability. 
Changes of over an order of magnitude can be seen at $\sim$2\,\kev, while at
$\sim$10\,\kev\ the variability is much more modest, at most a factor of
$\sim$2. The variability at lower energies ($\lesssim1$\,\kev) is also more
modest, although this is not surprising, as traditionally below $\sim$3\,\kev\
the spectrum observed from \ngc\ is dominated by diffuse thermal emission,
resolved with \chandra\ (\citealt{Wang09}), which should not vary on the
timescales probed. Variability of this nature and magnitude is most readily
explained with drastic changes in the line-of-sight absorption, which have
previously been observed from \ngc. This is confirmed by our detailed spectral
analysis, presented in Section \ref{sec_refl} (see also \citealt{Parker14}).
However, in past high S/N observations the column density has always remained
substantial, above $\sim$10$^{23}$ cm$^{-2}$ (\eg \citealt{Risaliti09a,
Maiolino10, Walton10Hex, Brenneman13}), occasionally increasing to the extent
of becoming Compton thick (\citealt{Risaliti05a}). Here, we are witnessing
variations in the opposite sense, observing the source in an almost fully
uncovered state for the first time.

The first of these four observations, which is also the most absorbed, displays
residuals to a simple absorbed powerlaw continuum indicating the presence
of relativistic disk reflection., \ie a relativistically broadened iron line at
$\sim$6\,\kev, and a prominent Compton hump at $\sim$30\,\kev\
(\citealt{Risaliti13nat}). Here, we apply the same continuum to the average
spectra from all four observations, consisting of a powerlaw with $\Gamma =
1.75$ modified by partially covering neutral absorption (using \pcfabs), with
both the column and the covering fraction free to vary between observations. 
This is fit to the 2.5--4, 7--10 and 50--80\,\kev\ energy ranges, in order to
fit the AGN dominated continuum, and avoid the most prominent reflection
features. Although formally only a visual aid, the residuals in the 3--80\,\kev\
energy range are extremely similar for each observation (see Figure
\ref{fig_ratio}, \textit{right panels}), despite the massive changes in absorption
column density inferred (ranging from $\sim$1--25$ \times
10^{22}$\,cm$^{-2}$). Relativistically broadened iron emission lines, extending
down to $\sim$4\,\kev, and strong Compton humps are observed in each case,
which are clearly independent of the line-of-sight absorption. These residuals
are not resolved by fitting this simple absorbed powerlaw model to the full
2.5--80\,\kev\ bandpass considered here.

In addition to the broad reflection residuals, two pairs of resolved narrow
absorption lines at $\sim$7 and 8\,\kev\ from ionised iron can also be seen in
all observations. These lines have been observed repeatedly from \ngc\ (\eg
\citealt{Risaliti05b, Brenneman13}), and will be studied in detail in subsequent
work (Nardini \etal, \textit{in prep}). However, although these features are not
the main focus of this work, we do point out that they primarily cause the sharp
drop in the spectrum just below $\sim$7\,\kev, the depth of which is essentially
independent of the neutral absorption column. This should not be confused
with either the iron edge associated with neutral absorption, or the blue wing of
the iron line. The blue wing of the line in fact continues to higher energies, and
can still be observed at $\sim$7.5\,\kev.

\subsection{Linking the Broad Iron Line and the Compton Hump}
\label{sec_EWR}

Although we show the time averaged spectra from each observation in Figure
\ref{fig_ratio} for clarity, \ngc\ is known to display both flux and spectral 
variability on relatively short timescales in soft X-rays ($\sim$10s of ks, \eg
\citealt{Risaliti09a}). As shown in Figure \ref{fig_lc}, such behaviour is clearly
displayed during our observations. Following \cite{Risaliti13nat}, we split each
observation into four separate periods (resulting in sixteen periods in total, as
highlighted in Figure \ref{fig_lc}), determined by changes in intrinsic flux, as
roughly indicated by the 5-10\,\kev\ count rate, and/or changes in spectral
shape, as roughly indicated by the hardness ratio between the 3--5 and
5--10\,\kev\ energy bands, and extracted spectra for each of these. For the
first observation, the selected periods are the same as those used in
\cite{Risaliti13nat}, for consistency. The data reduction follows that described
in section \ref{sec_red} for each period apart from the fourth period of
observation 3 (P12 in Figure \ref{fig_lc}, the brightest part of the least absorbed
observation). During this period, the observed \xmm\ countrates were formally
just in excess of the recommended limits for pile-up. Although comparison of
the pattern distribution suggests any pile-up effects below 10\,\kev\ are still
minimal, to be conservative we extract \xmm\ spectra using an annular region
for this period, with the same outer radius as the circular regions used
otherwise, but with the innermost 6$''$ of the PSF removed for \epicpn, and the
innermost 10$''$ removed for \epicmos.

Following the same procedure as above for each of these individual periods
results in the same residuals highlighted in Figure \ref{fig_ratio} \textit{in each
case} (albeit at lower S/N), even though the individual periods sample an even
broader range of absorption states (see below). In our further analysis, we only
consider the \xmm\ data above 2.5\,\kev\ in order to avoid the diffuse thermal
contribution and focus on these reflection features, and the \nustar\ data are
modeled over the 4--79\,\kev\ energy range. The only exception to this is
the \epicmos\ data from observation 3, in which narrow residuals are
observed just above 2.5\,\kev, most likely associated with calibration of the
instrumental edges at slightly lower energies; these data are therefore
modeled above 2.7\,\kev.

A key expectation of the relativistic reflection model is that the strength of the
iron line and the high energy reflection hump relative to the continuum are
positively correlated. In order to test this prediction in a model-independent
manner, we applied two purely phenomenological models to each of our 16
broadband spectra. The first consists of an absorbed powerlaw continuum, a
broad Gaussian emission line to treat the iron emission and a narrower
Gaussian absorption line to treat the strongest of the ionised iron absorption
features and avoid strong residuals which may alter the best fit values of the
other components, while the second consists simply of a powerlaw and a
neutral reflection continuum (\pexrav; \citealt{pexrav}). We simply use a single
Gaussian line to treat the ionised iron \ka\ absorption here as individually the
spectra from the 16 selected periods do not have sufficient S/N to separate
and constrain all four of the absorption lines visible in Figure \ref{fig_ratio}
simultaneously, and the two weaker iron \kb\ lines have a negligible effect on
this analysis. The first model is applied to the data below 10\,\kev, and the
second to the data above 10\,\kev, with only the parameters of the intrinsic
powerlaw continuum required to be the same in each energy range. In this way,
the two interesting components are treated completely independently. For
simplicity, we assume solar abundances for the reflection at this stage. The fits
obtained with this simple procedure give a reduced $\chi^2<1.2$ in all cases,
and do not show large residual features. We then estimated the equivalent
widths of the broad Gaussian components, and the ratio $R$ between the
reflection and intrinsic powerlaw normalizations.

\begin{figure}
\hspace*{-0.5cm}
\epsscale{1.12}
\plotone{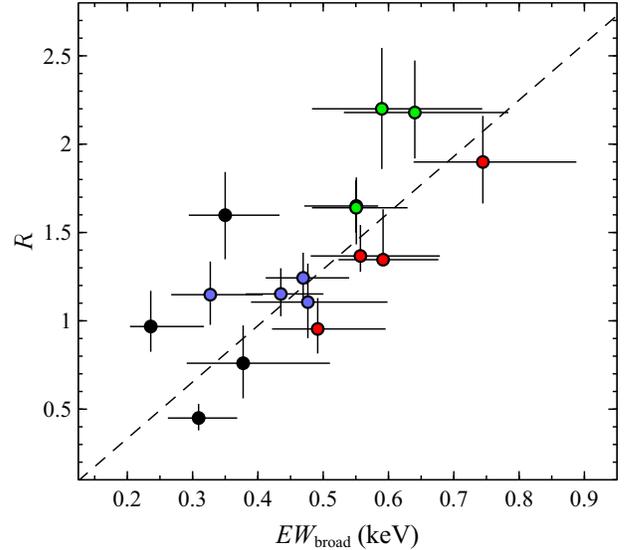}
\caption{
The correlation between the equivalent width ($EW_{\rm broad}$) of the broad
iron line, and the strength of the reflection continuum relative to the intrinsic
powerlaw continuum, quantified as the reflection parameter $R$, inferred from
the Compton hump for each of the 16 periods highlighted in Figure \ref{fig_lc}
using our simple phenomenological analysis. For clarity, the data from each
observation has been color-coded to match Figure \ref{fig_ratio}. The best fit
linear relation is shown with the dashed line.
}
\hspace*{0.5cm}
\label{fig_EWR}
\end{figure}

The results are plotted in Figure \ref{fig_EWR}. Although there is some scatter,
likely related to degeneracies arising from the simplistic analysis, a significant
positive correlation is still clear from the data, demonstrating a clear link
between the strength of the Compton hump and the strength of the iron
emission. The Kendall's $\tau$ correlation coefficient for these parameters is
$\tau=0.55$, implying the null hypothesis (no correlation) can be rejected at
just over 3$\sigma$ confidence. We stress again that we have endeavored to
treat the iron emission and the Compton hump independently here, and for
both features we consider quantities that assess their strength \textit{relative
to the continuum}, in order to ensure that our analysis would not artificially
produce such a correlation. If we fit a powerlaw relation to the data, the best
fit index, $1.55\pm0.36$ is consistent with the relationship being linear, as
would naturally be expected for two aspects of the same fundamental
emission component. For illustration, we show the best fit linear relation in
Figure \ref{fig_EWR}.

\begin{figure*}
\hspace*{-0.35cm}
\epsscale{0.54}
\plotone{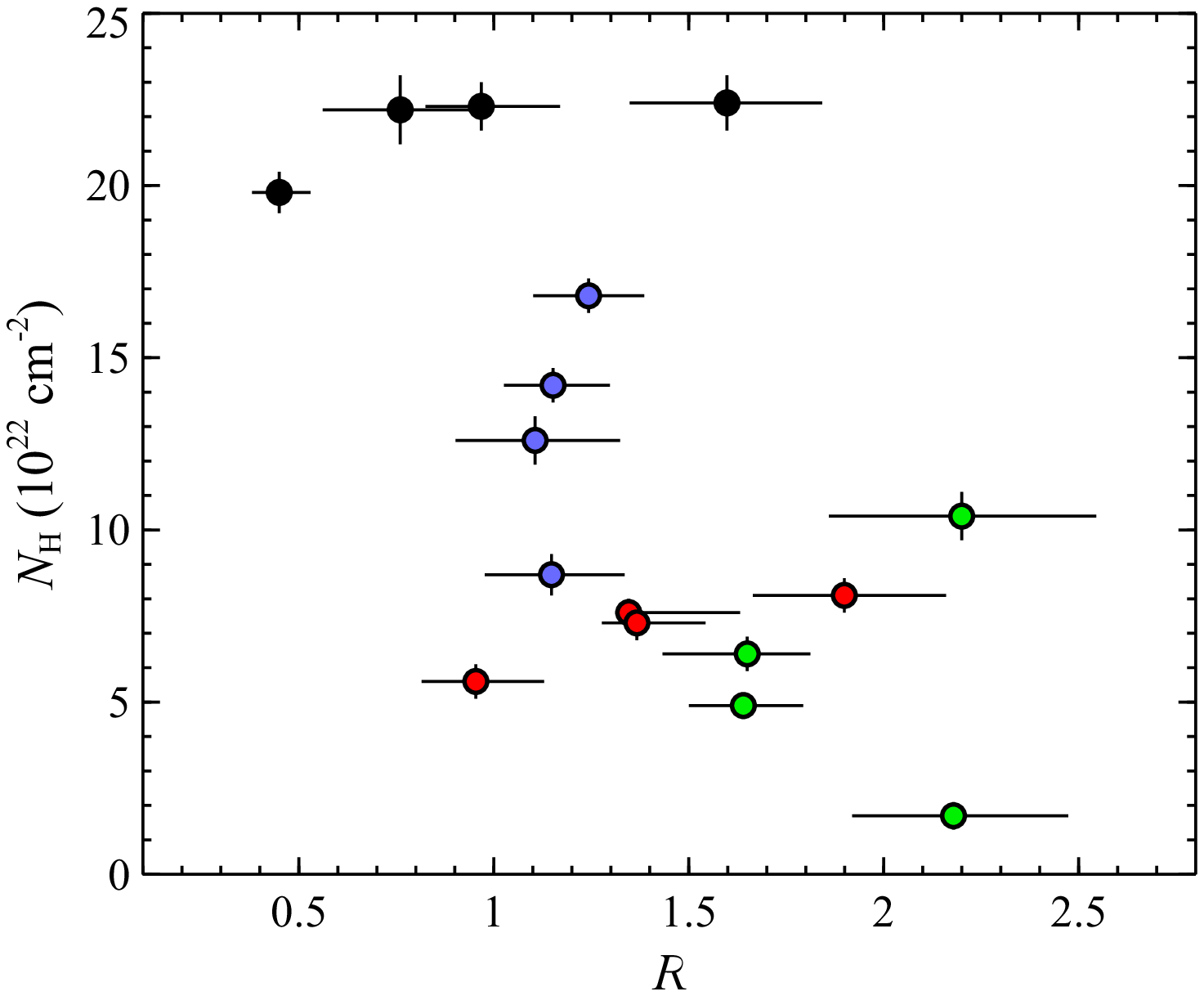}
\hspace{0.75cm}
\plotone{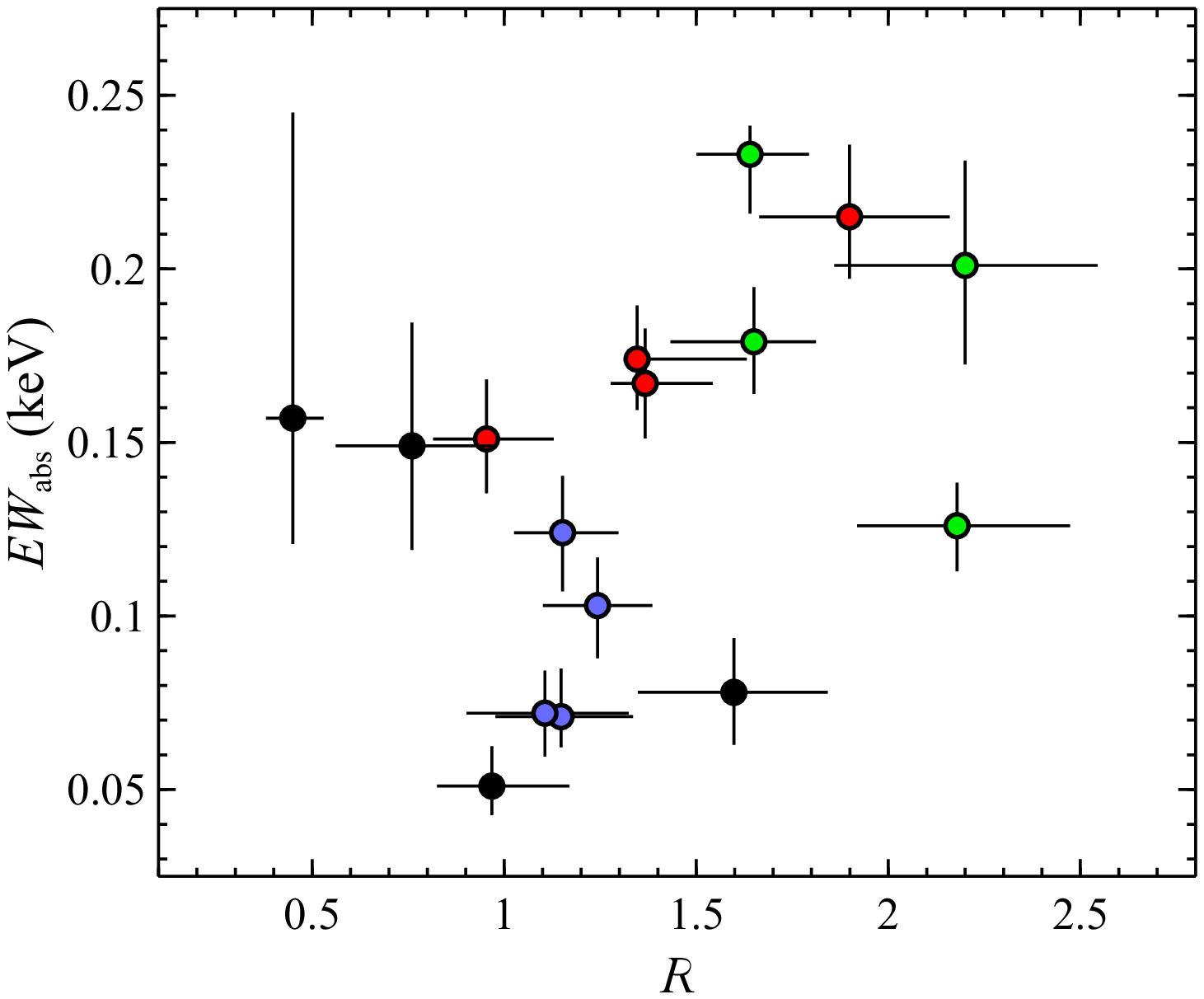}
\caption{
The reflection strength ($R$) plotted against the neutral column density (\nh)
and the inferred strength of the ionised iron absorption ($EW_{\rm abs}$)
obtained with our simple phenomenological analysis. The color-coding is as
for Figure \ref{fig_EWR}. In contrast to the strength of the broad iron emission,
neither of these quantities strongly correlate with $R$, implying the $EW_{\rm
broad}$--$R$ correlation shown in Figure \ref{fig_EWR} is intrinsic, and not merely
a result of parameter degeneracies arising from the nature of this analysis.
}
\hspace*{0.5cm}
\label{fig_Rvs}
\end{figure*}

Finally, we confirm the correlation observed between the broad iron emission
and the Compton hump is not itself an artificial result of any parameter
degeneracies related to this simple phenomenological analysis. Figure
\ref{fig_Rvs} shows the reflection parameter $R$ plotted against the column
density, and the equivalent widths of the Gaussian included to account for the
ionised absorption obtained with this analysis, which might influence the
strength inferred for the broad iron emission. In contrast to the broad Fe K
equivalent widths, no visible correlations are readily apparent between either
$R$ and \nh\ or $EW_{\rm abs}$. Applying the Kendall correlation test again,
we find the null hypothesis cannot be rejected at even the 95\% level in either
of these cases, confirming the intrinsic correlation between $R$ and $EW_{\rm
broad}$.

\subsection{Relativistic Disk Reflection}
\label{sec_refl}

Having demonstrated the link between the iron line and the Compton hump
using simple phenomenological models, we now undertake a detailed analysis
of the disk reflection utilizing a detailed, physically self-consistent reflection
model. We follow a similar approach to that undertaken in \cite{Risaliti13nat},
where we model each of the four periods identified from each observation
simultaneously, in order to account for the spectral variability. However, our
main aim in this work is to test the consistency of the results obtained with
the relativistic disk reflection interpretation over the vastly differing absorption
states, so we model each of the four observations independently.

\subsubsection{Model Setup}

In \cite{Risaliti13nat} we made use of the \reflionx\ reflection code
(\citealt{reflion}), however in this work we make use of the \xillver\ reflection
code (\citealt{xillver}) in order to test whether the results obtained previously
depend on the exact reflection model used. Similar to \reflionx, \xillver\
self-consistently incorporates both the reflected continuum, including the
high-energy ($\sim$30\,\kev) Compton reflection hump, and the
accompanying line emission. The version of \xillver\ used here also includes
the subtle effects of viewing angle on the intrinsic reflected emission, and
includes additional atomic physics. Besides the inclination ($i$), the other key
parameters are the iron abundance ($A_{\rm Fe}$) and the ionization state
(quantified as $\xi = 4\pi F/n$, where $F$ is the ionising flux at the surface of
the disk, and $n$ is the density) of the reflecting medium, and the photon
index of the ionizing continuum (assumed to be a powerlaw). To fully model the
relativistic disk reflection, the \xillver\ model is convolved with the \relconv\
kernel to account for the relativistic effects close to a black hole
(\citealt{relconv}). The key parameters for \relconv\ are the spin of the black
hole, $a^*$, the inclination of the accretion disk, and the index of the radial
emissivity profile of the reflected  emission, $q$, assumed here to have a
powerlaw form. In order to measure the spin, we assume the accretion disk
extends down to the ISCO; should this not be the case, any spin measurement
obtained instead becomes a lower limit.

In addition to the disk reflection, we include an intrinsic powerlaw continuum.
Both of these emission components are modified with a partially covering
neutral absorber, and four multiplicative Gaussian absorption lines (using
{\small GABS}) in order to account for the known ionised absorption visible in
Figure \ref{fig_ratio} (with the higher S/N in the combined data considered
during this analysis, all four lines can now be independently constrained). We
use simple Gaussian absorption lines in this work in order to also test whether
the spin measurement depends strongly on our prior treatment of the ionised
absorption with the \xstar\ photoionisation code (\citealt{xstar}). Finally, we
also include a distant, neutral reflection component (again using \xillver) in
order to model the narrow component of the iron line, but this is not modified
by either of the absorption components highlighted above, as this may well be
reprocessed emission from the neutral absorber itself, which seems to be
associated with the broad line region (\citealt{Risaliti09a}), or could also
plausibly arise from even more distant material, e.g. the narrow-line region
and/or the dusty torus. For self-consistency, all inclination parameters, iron
abundance parameters and photon indices are linked between the various
components included in the model, and each of the four Gaussian absorption
lines are required to have the same relative line broadening and outflow
velocity.

As stated previously, for each observation we analyze the four periods
identified in Figure \ref{fig_lc} simultaneously. All parameters that are not
expected to vary on observable timescales, i.e. the black hole spin, accretion
disk inclination, the system iron abundance and the normalization of the
distant neutral reflection, are linked between the four periods, but allowed to
vary overall. Following \cite{Risaliti13nat}, we also assume that the photon
index remains constant. However, the neutral absorption column and the
relative contribution of the intrinsic continuum and the disk reflection are
allowed to vary between the periods. We also investigated allowing the
covering fraction of the neutral absorber and the emissivity profile of the
reflected emission to vary, but we found that doing so did not result in a
significant statistical improvement for any of the observations. These
parameters are therefore also linked between the periods, but allowed to
vary overall. We also initially allowed the ionised absorption lines to vary, but
found that this was only statistically required in the third observation (January
2013). In this case, we allow the line strengths to vary between the selected
periods, but assume the relative strengths of the associated \ka\ and \kb\
lines, the line broadening and the outflow velocity do not vary with time,
although they are allowed to vary overall. For the other observations, the line
strengths, widths and the outflow velocities are assumed constant with time,
but are again free to vary overall.

Since the intrinsic flux changes are only moderate within each individual
observation (see the 5--10\,\kev\ lightcurves in Figure \ref{fig_lc}), we
assume the disk ionization does not vary between the selected periods, and
we also limit the ionization to the range $\log\xi_{\rm disk}<3$. We do this
to exclude unphysical reflection-dominated solutions at low spin, which
require an extremely ionised accretion disk ($\log\xi_{\rm disk} \geq 3.5$)
and a very hard intrinsic powerlaw continuum ($\Gamma \sim 1.5$), only
contributing significantly at the highest energies probed ($E \gtrsim
50$\,\kev). Such scenarios can be rejected in the standard disk reflection
paradigm, since the enhanced lightbending required to produce
reflection-dominated spectra in turn requires the corona to be located within
a few gravitational radii (\rg) of the black hole, which is only plausible when
the spin of the black hole is very high (\citealt{lightbending}). However, this
is not yet self-consistently incorporated into the disk reflection models
currently available, so such solutions must be excluded manually. For
comparison, we note that such solutions had previously been allowed when
computing the spin  constraint presented in \cite{Risaliti13nat}. In any case,
the ionization of the disk is only expected to be moderate, based on results
obtained for other AGN ($\log\xi_{\rm disk} < 3$; \citealt{Ballantyne11,
Walton13spin}). As a final technical detail, we also assume that the
cross-calibration constants between the various detectors utilized in this
work do not vary with time, apart from during observation 3, owing to the
unique data reduction procedure for period 12 (see section \ref{sec_EWR}).
In this case, only the first 3 periods (9, 10 and 11) are required to have the
same cross-calibration constants.

\begin{table}
  \caption{Key continuum parameters obtained for each of the four
observations of \ngc.}
\begin{center}
\begin{tabular}{c c c c c c}
\hline
\hline
\\[-0.2cm]
Parameter & \multicolumn{4}{c}{Observation} \\
\\[-0.275cm]
& 1 & 2 & 3 & 4 \\
\\[-0.3cm]
\hline
\hline
\\[-0.15cm]
$\Gamma$ & $1.84^{+0.02}_{-0.04}$ & $1.99^{+0.04}_{-0.20}$ & $2.05^{+0.05}_{-0.03}$ & $2.07^{+0.04}_{-0.05}$ \\
\\[-0.25cm]
\nh(1)\tmark[a] & $25.2\pm0.7$ & $7.7^{+0.5}_{-0.8}$ & $12.5^{+1.2}_{-1.3}$ & $7.7^{+0.3}_{-0.4}$ \\
\\[-0.25cm]
\nh(2)\tmark[a] & $21.8\pm0.6$ & $7.4^{+0.5}_{-0.8}$ & $6.3^{+1.2}_{-0.8}$ & $12.7^{+0.8}_{-0.6}$ \\
\\[-0.25cm]
\nh(3)\tmark[a] & $26.4\pm0.9$ & $7.1\pm0.5$ & $2.9\pm0.4$ & $14.0\pm0.5$ \\
\\[-0.25cm]
\nh(4)\tmark[a] & $22.5\pm0.9$ & $6.3^{+0.4}_{-0.8}$ & $1.3^{+0.3}_{-0.2}$ & $18.3\pm0.6$ \\
\\[-0.25cm]
$C_{\rm f}$ (\%) & $97.2\pm0.2$ & $88^{+1}_{-2}$ & $80^{+3}_{-4}$ & $95.5^{+0.3}_{-0.4}$ \\
\\[-0.25cm]
$\log\xi_{\rm disk}$\tmark[b] & $1.6^{+0.2}_{-1.5}$ & $>1.5$ & $1.7^{+0.1}_{-1.2}$ & $1.7^{+0.1}_{-1.1}$ \\
\\[-0.25cm]
$q$ & $7.2^{+0.5}_{-1.8}$ & $6.5^{+0.2}_{-0.5}$ & $8.3^{+0.2}_{-2.1}$ & $5.0^{+0.5}_{-0.4}$ \\
\\[-0.25cm]
$A_{\rm Fe}/\rm{solar}$ & $5.0^{+2.3}_{-0.7}$ & $5.0^{+1.4}_{-0.8}$ & $5.0^{+1.0}_{-0.2}$ & $3.8^{+0.4}_{-0.3}$ \\
\\[-0.25cm]
\hline
\\[-0.15cm]
\multirow{2}{*}{\chisq/DoF} & 3837/3747 & 6801/6569 & 6067/5783 & 5645/5689 \\
& (= 1.02) & (= 1.04) & (= 1.05) & (= 0.99) \\
\\[-0.2cm]
\hline
\hline
\\[-0.25cm]
\end{tabular}
\end{center}
$^{a}$ Column densities 1--4 refer to the 4 periods selected within each respective
observation (\eg \nh(3) for observation 4 refers to P15 in Fig \ref{fig_lc}), and are
quoted in units of $10^{22}$\,\atpcm. \\
$^{b}$ The ionization parameter is quoted in units of \ergcmps. \\
\label{tab_continuum}
\end{table}

\begin{figure*}
\hspace*{-0.6cm}
\epsscale{1.17}
\plotone{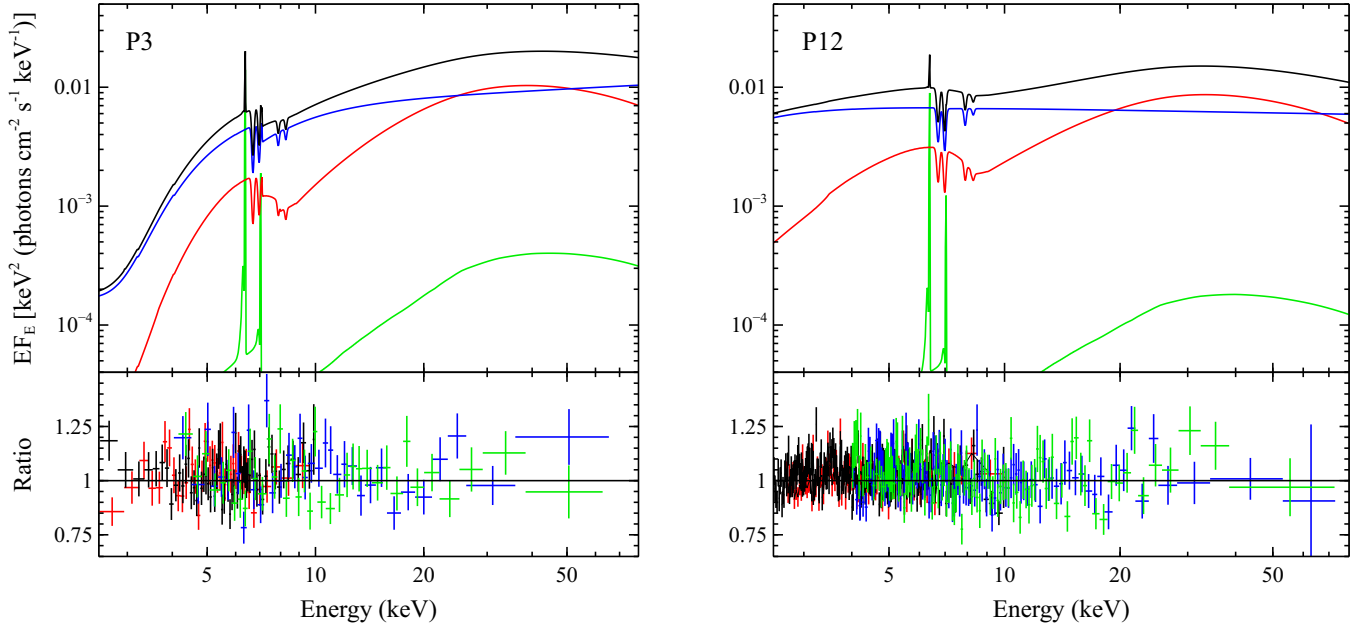}
\caption{
The relativistic reflection model applied to two of the selected periods (see
Figure \ref{fig_lc}). For illustration, we show the most (P3) and the least
absorbed (P12) periods. \textit{Top panels:} the best fit models obtained for
the two periods. In each case, the total model is shown in black, the
powerlaw continuum in blue, inner disk reflection in red and distant
reflection in green. \textit{Bottom panels:} data/model ratios for the best fit
reflection models for these two periods. The \epicpn, \epicmos\ FPMA and
FPMB data are shown in black, red, green and blue, respectively.
}
\hspace*{0.25cm}
\label{fig_fit}
\end{figure*}

\subsubsection{Results}

We systematically apply this modeling procedure to each of the four coordinated
observations independently, \ie although we require many of the parameters to
be the same for each of the periods identified within individual observations (as
detailed above, see Figure \ref{fig_lc}), no requirement for \textit{any parameter}
to be the same between observations is imposed. This allows us to test the
consistency of the results obtained from each of the observations for key physical
parameters, specifically the disk inclination and the black hole spin (which should
not change on the timescales probed), and therefore to test the robustness of the
relativistic disk reflection interpretation with a consistent modeling procedure.
With this approach, we obtained good fits for each observation, \chisq/DoF =
3837/3747 (= 1.02), 6801/6569 (= 1.04), 6067/5783 (= 1.05) and 5645/5689
(= 0.99) respectively, and no obvious systematic residuals remain. We show the
fits to two of the selected periods in Figure \ref{fig_fit}, for illustration, and the
key continuum parameters obtained are presented in Table \ref{tab_continuum}
(with the exception of the black hole spin and disk inclination which, as the main
focus of this work, are presented separately in Table \ref{tab_spin}).

\begin{figure*}
\hspace*{-0.6cm}
\epsscale{1.16}
\plotone{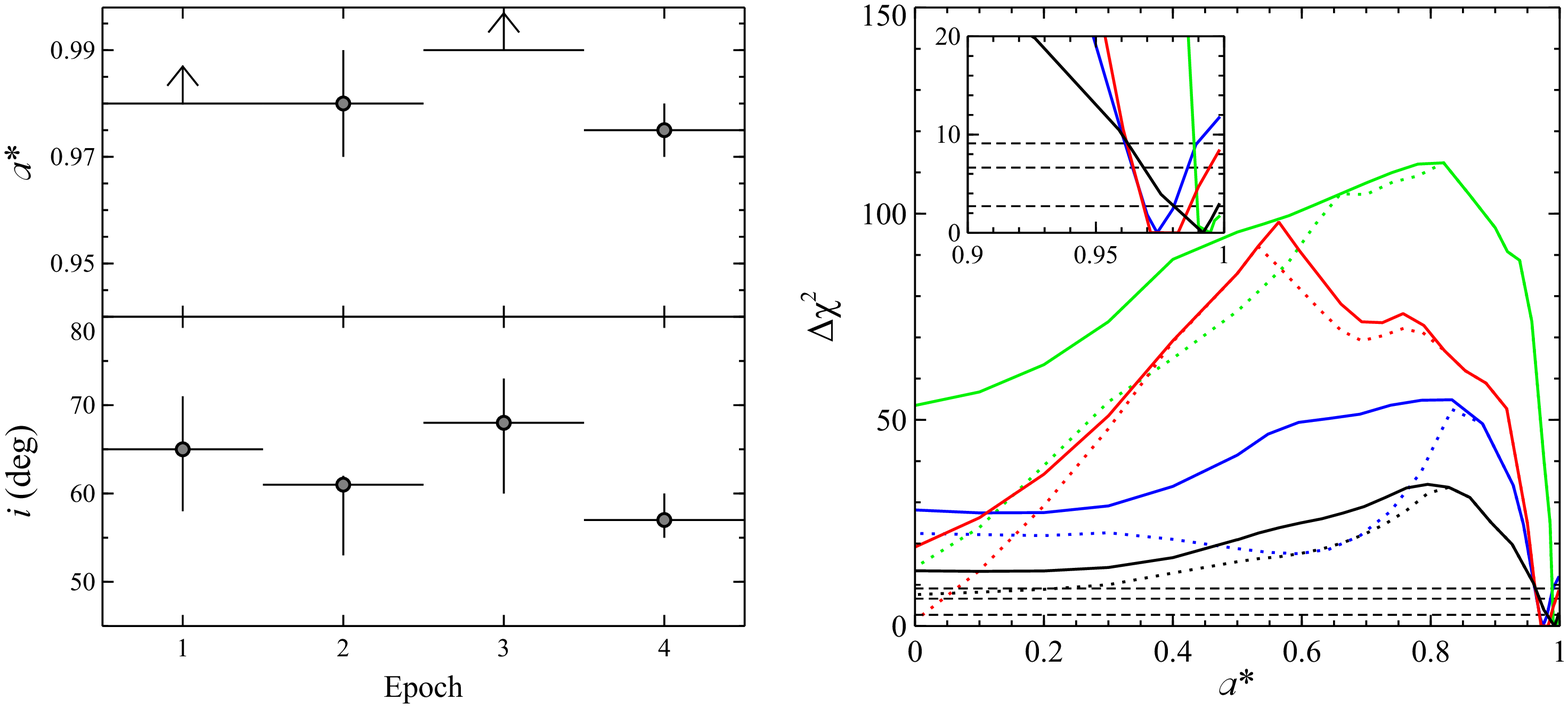}
\caption{
\textit{Left panel:} The results obtained for the black hole spin (\textit{top}) and
the disk inclination (\textit{bottom}) with our time-resolved analysis of each of
the four coordinated \nustar+\xmm\ observations. The errors are the 90\% 
confidence intervals for a single parameter of interest. \textit{Right panel:} The
$\Delta$\chisq\ confidence contours for the black hole spin. As in Figure
\ref{fig_ratio}, the contours for observations 1, 2, 3 and 4 are shown in black, red,
green and blue respectively. The horizontal dashed lines represent the 90, 95
and 99.97\% (\ie 3$\sigma$) confidence levels for a single parameter of interest.
Solid contours are calculated limiting $q<3.5$ for $a^* \lesssim 0.85$, as
expected for the standard disk reflection/lightbending model, while the dotted
segments are calculated relaxing this constraint.
}
\hspace*{0.25cm}
\label{fig_spin}
\end{figure*}

In Figure \ref{fig_spin} (\textit{left panel}) we show the spin and inclination
constraints obtained from each of the four observations. It is clear that in each
case the spin is inferred to be very high ($a^* > 0.97$), and the inclination is
also inferred to be fairly high ($i \sim 60$\deg). The quantitative constraints
obtained are presented in Table \ref{tab_spin}. The quoted uncertainties
represent only the statistical uncertainties obtained, and do not incorporate any
systematic uncertainties associated with the modeling procedure itself, or the
models used. For example, although we have not allowed such solutions in order
to enforce some level of physical self-consistency at low spin, some of the 
high-ionisation, reflection-dominated solutions might be considered physically
plausible at high spins, and permitting these solutions could broaden the
statistically allowed parameter range to some extent. However, we stress that,
based on the physical considerations discussed above, this would not change
the ultimate requirement for high spin. Furthermore, we note that when the spin
constraints have been investigated for a variety of input model assumptions for
other accreting black holes (both binaries and AGN) in which high spin has been
inferred, the overall requirement for high spin has generally been found to be
robust (\eg AGN: \citealt{Brenneman11}; Galactic binaries: \citealt{Reis1836,
Walton12xrbAGN, Tomsick14}). 

In addition, although both the reflection models utilized here and in
\cite{Risaliti13nat} compute reflected spectra from simple, constant-density
slabs (\citealt{reflion, xillver}), models produced with more complex density
profiles which might be expected for real accretion disks -- \eg models
computed from slabs of material in hydrostatic equilibrium
(\citealt{Nayakshin00}) -- also produce broadly similar results, \ie strong iron
emission that is narrow in the restframe, and a prominent Compton hump.
Thus, there is no reason to expect that such models would produce noticeably
different results to those obtained here, as the line broadening will still
naturally be dominated by the inner extent of the disk. Indeed,
\cite{Ballantyne01, Ballantyne04} demonstrate the qualitative similarity of
models assuming a constant density and those invoking more complex
density profiles across the bandpass incorporating the iron \ka\ emisison
and the Compton reflection hump, further supporting our assertion that the
key results presented here regarding the black hole spin and disk inclination
are robust to these particular details. Instead, we expect that the iron
abundance might be the parameter most sensitive to the assumptions
regarding the density profile as, for certain illumination scenarios, models with
different density treatments can make different predictions for the strength of
the iron emission relative to the continuum (\citealt{Nayakshin00}).

The quoted uncertainties also do not incorporate systematic uncertainties
related to the key assumptions inherent to the measurement of black hole spin,
most importantly that the accretion disk truncates at the ISCO (or more precisely
that there is negligible reflected emission from within the ISCO). However,
simulations suggest this is generally a fairly good assumption, particularly at
high spin (\citealt{Reynolds08}). Nevertheless, even with these formal statistical
uncertainties, both the spin and the inclination obtained independently from
each observation all agree at the 3$\sigma$ level of confidence or better, clearly
demonstrating that the reflection interpretation naturally gives consistent results
for the key physical quantities, despite the vast range of absorption states
observed. A full assessment of the model/systematic uncertainties when
self-consistently treating all the data will be presented in future work
(Brenneman et al. \textit{in preparation}). In addition to the key geometric
parameters, the iron abundance should also remain constant on observable
timescales, and again the values obtained display a good level of consistency,
agreeing at the 3$\sigma$ level or better.

\begin{table}
  \caption{The black hole spin and disk inclination constraints obtained for each
of the four observations of \ngc.}
\begin{center}
\begin{tabular}{c c c c c c c}
\hline
\hline
\\[-0.2cm]
& Observation & & $a^*$ & & $i$ (deg) & \\
\\[-0.25cm]
\hline
\hline
\\[-0.15cm]
& 1 & & $>0.98$ & & $65^{+6}_{-7}$ \\
\\[-0.25cm]
& 2 & & $0.98 \pm 0.01$ & & $61^{+1}_{-8}$ \\
\\[-0.25cm]
& 3 & & $>0.99$ & & $68^{+5}_{-8}$ \\
\\[-0.25cm]
& 4 & & $0.975 \pm 0.005$ & & $57^{+3}_{-2}$ \\
\\[-0.25cm]
\hline
\hline
\\[-0.15cm]
\end{tabular}
\label{tab_spin}
\end{center}
\end{table}

We also show the $\Delta$\chisq\ confidence contours for the spin parameter
obtained from each of the observations in Figure \ref{fig_spin} (\textit{right 
panel}) in order to demonstrate the best fit solutions obtained are always
consistent with being the global minimum when considering prograde 
spin\footnote{We do not consider retrograde spin in this work, as to date there
is no evidence for an active galaxy with such a spin orientation
(\citealt{Walton13spin}); while \cite{Cowperthwaite12} initially suggested that
3C\,120 might have a retrograde spin, \cite{Lohfink13} later demonstrated this
is unlikely to be the case.}, despite the moderately complex \chisq\ landscape.
Similar to recent analyses by \cite{Steiner12lmc} and \cite{King14}, when
computing these contours we restrict the emissivity index to $q<3.5$ for $a^*
\lesssim 0.85$, as solutions with steep emissivities are only physically plausible
for rapidly rotating black holes in the disk reflection/lightbending paradigm
(\eg \citealt{Wilkins12, Dauser13}). Although something of an arbitrary value,
we make this transition at $a^* \sim 0.85$ for two reasons, first because this is
always a low enough spin that it doesn't influence the statistical constraint
obtained around the best-fit solution, and second to avoid large discontinuities
in the plotted contours, since the best fit models for each of the four
observations all give $q \lesssim 3.5$ at $a^* \sim 0.85$. While each individual
case is slightly different, in general, the \chisq\ contours obtained from each
observation with the above model setup favor a high spin, with a rapid initial
rise in \chisq\ followed by a slow decline as lower spins are considered.

Finally, we also demonstrate the effect of relaxing the constraint on the
emissivity index for low spins (dotted contour segments). This generally results
in slightly better fits in this regime, although the high spin solution is still
statistically favoured for most of the observations. In the case of observation 2,
however, the decrease in \chisq\ towards low spin becomes such that the
best fit at $a^*$ = 0 is formally statistically comparable to the best-fit high spin
solution, owing to strong degeneracies among the various model parameters
(when they are allowed sufficient freedom). Interestingly, this is the
observation displaying the least spectral variability, which may in fact serve to
exacerbate parameter degeneracies for the complex model considered here.
Evidence for this can also be seen in Table \ref{tab_continuum}, where several
key continuum parameters are not as well constrained as for the other
observations. However, in addition to requiring a steep ($q>4$) emissivity
index, this low spin solution is also reflection-dominated, despite our best
efforts to exclude such solutions in this regime. We therefore discard this
low-spin solution on physical grounds. Models that self-consistently
incorporate spin-dependent effects for the emissivity and the reflected
contribution (\eg Dovciak et al, \textit{in preparation}) will be utilized in the
future.

\section{Discussion}
\label{sec_dis}

We have presented a spectral analysis of the full set of four coordinated \nustar\
and \xmm\ observations of the well known Seyfert galaxy \ngc, focusing
primarily on the 2.5--79\,\kev\ energy range in which the nuclear emission
dominates. Although \ngc\ has typically been observed to be obscured by a
fairly substantial column (\eg \citealt{Risaliti05b, Walton10Hex, Brenneman13}),
as shown in Figure \ref{fig_ratio}, these four high S/N observations caught the
source in an unprecedented range of absorption states, including states with
little line-of-sight absorption to the central nucleus ($N_{\rm H} < 10^{22.5}$
cm$^{-2}$). The full implications for the structure of the absorption will be
explored in future work (Rivers et al. 2014, \textit{in preparation}). Despite this
range of absorption, the observed spectrum from each observation displays the
same hallmarks of reflection from the innermost regions of the accretion flow,
\ie relativistically broadened iron emission and a strong Compton hump.
Furthermore, we also find that the breadth of the line is independent of the
level of absorption observed. The simplest interpretation is that these features
are not directly associated with line-of-sight absorption, providing yet more
compelling evidence for the contribution of relativistic disk reflection.

Through time-resolved spectroscopy, similar to recent analyses of other bright
AGN (\eg \citealt{Reis3783, Parker14mcg, Marinucci14mcg6}), we have
demonstrated a clear link between the broad iron emission and the Compton
hump (Figure \ref{fig_EWR}), as expected if these features are two aspects of the
same underlying component, as in the relativistic reflection model. We also
examined the consistency of the results obtained for key physical parameters
through self-consistent modeling of this reflection, allowing for variable,
partially-covering absorption. In spite of the complexity of the behaviour
displayed by \ngc, consistent results are obtained for both the disk inclination
and the black hole spin in each case, as shown in both Figure \ref{fig_spin} and
Table \ref{tab_spin}. The statistical agreement between the results obtained is
good even before considering the potential systematic uncertainties associated
with these measurements. This is an important consistency check for the disk
reflection interpretation, since these physical parameters should not vary on
observable timescales. The fact that similar results are naturally obtained from
our reflection analysis provides further confidence in the results obtained, and
thus in the overall reflection paradigm, our ability to view the innermost regions
of the accretion flow and to measure black hole spin for active galaxies.

\subsection{Distant Reprocessing}

In response to the results presented by \cite{Risaliti13nat}, \cite{LMiller13}
performed a series of simulations arguing that a combination of partially
covering absorption and distant reprocessing could produce spectra similar to
those observed without any contribution from relativistic disk reflection. A
similar model had been investigated by \cite{Risaliti13nat}, and was found to
require three independent absorption components, one of which was
Compton-thick, in addition to Compton reflection from distant material in order
to reproduce both the $\sim$4--8\,\kev\ spectral complexity (otherwise
interpreted as relativistically broadened iron emission) and the strong
hard-excess at $\sim$20\,\kev. This model was able to adequately reproduce
the data from the first observation in a statistical sense, and we confirm that
this formally remains the case for the additional observations by repeating the
time-resolved analysis undertaken in \cite{Risaliti13nat} for each, constructing
models in which the relativistic disk reflection contribution is replaced by
additional partially covering absorption components.

Consistent with \cite{Risaliti13nat}, we find that two additional absorption
components (making three in total) are generally required to fit the data
(\chisq/DoF = 3853/3721, 6940/6544, 6093/5763 and 5664/5664 for the
four observations, respectively), in addition to a distant Compton reflection
component to account for the narrow iron line (modeled again with \xillver).
These fits are still poorer than for the relativistic reflection model (see Table
\ref{tab_continuum}). Removal of one of these absorption components
degrades the fit by $\Delta\chi^{2}$ = 30--70 (for eight fewer free
parameters) in each of the observations, with the third absorber being least
statistically significant in the least absorbed observation (Obs. 3, although an
F-test still suggests it is significant at greater than the 99.9\% statistical level).
The column densities and covering fractions for the two additional absorbing
components are found to range from $N_{\rm{H,1}} = 3.5-5\times10^{24}$
cm$^{-2}$ and $C_{\rm{f,1}} = 0.5-0.7$ (Compton-thick), and $N_{\rm{H,2}}
= 1-5\times10^{23}$~cm$^{-2}$ and $C_{\rm{f,2}} = 0.4-0.8$ respectively,
while the third component displays very similar values to those obtained with
the relativistic reflection model (Table \ref{tab_continuum}), again displaying
more than an order of magnitude of variation in column density. Regardless
of whether two or three absorption components are used, one is strongly
required to be Compton-thick and relatively constant, and one is strongly
required to be highly variable. However, this model was ultimately rejected on
physical grounds in \cite{Risaliti13nat}. We summarize the main points below,
and refer the reader to that work for a more detailed discussion.

\begin{table}
  \caption{Relative and absolute line fluxes for the narrow iron emission for each
of the four observations of \ngc.}
\begin{center}
\begin{tabular}{c c c c c c c}
\hline
\hline
\\[-0.2cm]
& \multicolumn{5}{c}{Observation} \\
& 1 & 2 & 3 & 4 & \\
\\[-0.3cm]
\hline
\hline
\\[-0.15cm]
$EW_{\rm narrow}$ (eV) & $57^{+12}_{-11}$ & $31\pm7$ & $31\pm8$ & $34^{+8}_{-7}$ \\
\\[-0.2cm]
$N_{\rm narrow}$\tmark[a] & $1.1^{+0.1}_{-0.2}$ & $1.2^{+0.3}_{-0.2}$ & $1.1^{+0.3}_{-0.2}$ & $1.0\pm0.2$ \\
\\[-0.25cm]
\hline
\hline
\\[-0.25cm]
\end{tabular}
\end{center}
$^{a}$ Narrow line normalisations, quoted in units of $10^{-5}$ \xspnorm\
(at 1\,\kev).
\label{tab_EW}
\end{table}

In the absorption dominated scenario, the intrinsic luminosities inferred are
significantly higher than the observed ones, due to the effects of Compton
scattering. The exact correction factor depends primarily on the geometry and
global covering fraction (as seen from the X-ray source; $C_{\rm g}$) of the
dominant Compton thick absorber, and may vary from a factor of $\sim$4-6
for an almost totally covering absorber (\ie large $C_{\rm g}$, in good
agreement with \citealt{LMiller13} despite the modeling differences), to
$\sim$20-30 for a geometrically thin gas ring (small $C_{\rm g}$). In all
scenarios, the intrinsic luminosities are either inconsistent with those estimated
from other indicators (smaller $C_{\rm g}$; in the thin-ring case, the total X-ray
luminosity alone is close to the Eddington limit, and more than one order of
magnitude higher than the luminosity inferred from the [OIII] emission) or the
observed level of reprocessed  radiation (\eg the narrow iron K$\alpha$ flux,
total radiation reprocessed in the infrared; larger $C_{\rm g}$).

\cite{LMiller13} argue in the higher $C_{\rm g}$ case the intrinsic X-ray flux in
the absorption scenario may be roughly similar to that inferred in the disk
reflection scenario, given that the lightbending required to explain the enhanced
reflection and steep emissivity profiles likely mean that the fraction of the
intrinsic X-ray flux lost over the event horizon is enhanced. However, the key
point is that in the absorption scenario this flux does escape from the black hole,
and must be reprocessed. Furthermore, though \cite{LMiller13} suggest almost
\textit{all} AGN (including type-1 sources) are affected by Compton-thick
line-of-sight obscuration in X-rays, with covering factors close to unity (see
also \citealt{Tatum13}) and highlight that \ngc\ does not display extreme levels
of reprocessing relative to other AGN, such comparison does not address the
original point from \cite{Risaliti13nat} that the observed level of reprocessing
does not appear to match that expected from the absorption scenario in an
\textit{absolute} sense, not a relative sense. Therefore, the basic conclusion
presented in \cite{Risaliti13nat} remains valid.

\subsection{Viewing the Inner Disk}

In addition to these considerations, the observed spectral variability from the
full dataset also allows us to confirm the contribution of relativistic reflection
from the inner disk through other physical arguments. Were the feature
interpreted as the iron line to be produced primarily by partially covering
line-of-sight absorption, rather than by disk reflection, there is no reason to
expect \textit{a priori} that such consistency in the inner disk parameters
would be observed. Indeed, given the strong evolution clearly observed in the
line-of-sight absorption, this would certainly not be expected, particularly for
the black hole spin which is primarily constrained by the red-wing of the line,
where the observed absorption variability has the largest effect (See Figure
\ref{fig_ratio}). 

Furthermore, as is also visually apparent from Figure \ref{fig_ratio}, the strength
of the narrow iron emission, which likely does arise through distant reprocessing,
varies relative to that of the broad iron emission. To quantify this, we construct
a simple phenomenological model, replacing the distant and disk reflection
components included in the self-consistent model used above (section
\ref{sec_refl}) with a Gaussian emission line (with an intrinsic width of 10\,eV)
and a combination of a reflected continuum (\pexrav) and a relativistic emission
line (\relline), and apply this model to the time-averaged broadband spectrum
from each of the observations. The equivalent widths of the narrow emission
line are presented in Table \ref{tab_EW}, confirming the variation in its strength
relative to the broad line. This strongly suggests that these features are not
different aspects of the same underlying emission component, particularly given
that the \textit{absolute} flux of the narrow line is consistent with remaining
constant over the observed timescales (Table \ref{tab_EW}), as expected for
distant emission.

Finally, we note that the detection of a short-timescale reverberation lag
associated with the broad emission line further confirms this as emission from
the inner regions of the accretion flow (Kara et al. 2014, \textit{in preparation}).
Observation of similar short-timescale reverberation from features potentially
associated with reflection from the inner disk has recently been found to be
fairly common, both from the soft excess (\citealt{FabZog09, Emman11b,
deMarco13, Fabian13iras, Cackett13}) and the broad iron line (\citealt{Zoghbi12,
Zoghbi13a, Kara13feK}), also strongly supporting an origin in the inner disk. 
Although reverberation from distant reprocessing material was again initially
proposed as an alternative interpretation for the early AGN lag results (\eg
\citealt{LMiller10b, Legg12}), this no longer seems able to explain the complex
lag phenomenology observed. In particular, high-frequency reverberation lags
have now been detected from the broad iron \ka\ emission (\eg
\citealt{Zoghbi12, Kara13feK}), while similar iron \ka\ features are not seen in
the low-frequency lags relevant to the distant reverberation interpretation (e.g.
\citealt{Kara13feK}), and the association of these low-frequency `hard' lags
with the powerlaw continuum, rather than distant reverberation, has now been
demonstrated (\citealt{Walton13lag}), a conclusion supported by comparison
with Galactic binaries (\citealt{Kotov01}). 

Indeed, the similarity of the broad iron features observed from both AGN and
Galactic BHBs, which are generally not heavily absorbed, in itself provides a
compelling case for a disk reflection origin (\citealt{Walton12xrbAGN}),
and features consistent with relativistic disk reflection are also observed in AGN
with no evidence for obscuration (\eg \citealt{Nardini11, Lohfink12,
Walton13spin}). Furthermore, we have now detected relativistically broadened
iron emission in a gravitationally lensed quasar (\citealt{Reis14nat}), where
microlensing studies independently constrain the size of the X-ray emitting
region to be $R_{\rm X} \lesssim 10$\,\rg\ (\citealt{Dai10}), firmly ruling out
the possibility of the broadened iron emission being produced by large, distant
reprocessing structures. There is, therefore, now a substantial body of
independent evidence suggesting that we are indeed viewing the inner disk in a
significant population of AGN. Our results firmly place \ngc\ among this number
(with the obvious exception of the Compton-thick phases occasionally displayed;
\citealt{Risaliti05a}).

\subsection{Black Hole Spin}

These results present a consistent picture, that even though \ngc\ does clearly
exhibit variable line-of-sight absorption, it also hosts a rapidly rotating black
hole, viewed at moderately high inclination. Our analysis constrains the spin of
\ngc\ to be $a^* >0.97$ (based on the 90\% statistical uncertainties). The
quantitative results obtained for the spin here (Table \ref{tab_spin}) are
consistent with the initial constraint presented in \cite{Risaliti13nat} based on
the first observation alone, despite subtle differences in our analysis, \eg using
a different reflection model and a different treatment for the ionised iron
absorption, suggesting that the overall conclusion is not sensitive to these
details. The spin constraints obtained here are also similar to the constraint
obtained by \cite{Brenneman13}, and that implied from the inner radius obtained
by \cite{Walton10Hex} through study of a single \suzaku\ observation, although
in the former the inclination was not allowed to vary (we note that the assumed
value was similar to those obtained here), and the latter simply assumed fully
covering neutral absorption. Nevertheless, our results attest positively to the
broad reliability of single-epoch spin measurements from reflection for active
galaxies.

This is an important point, as it is difficult to collect the wealth of high-quality
data across various epochs that can be available for Galactic binaries (\eg
\citealt{Miller12}) for active galaxies, owing to both their longer variability
timescales and their typically lower fluxes. In many cases, studies of AGN spin
are therefore limited to single-epoch reflection measurements (\eg
\citealt{Walton13spin}). Indeed, in order for AGN spin measurements to progress
towards ultimately providing strong tests of models for galaxy formation and/or
SMBH growth (\eg \citealt{Berti08, Dubois13}), substantially larger and better
statistically defined samples of AGN spin measurements than are currently
available (\citealt{Walton13spin, Reynolds13rev}) will be required, covering a
broader range of redshift space, and hence pragmatically the reliance on such
single-epoch reflection measurements will inevitably increase\footnote{We
note that in rare cases in the local universe it may still be possible to directly
observe the thermal emission from the inner disk, from which it may be possible
to constrain the spin through estimation of the emitting area, provided the black
hole mass, disk inclination and the relevant atmospheric corrections are known
(\citealt{Zhang97, Done13}). However, given the disk temperatures involved for
even the lowest mass AGN, even the small likelihood of being able to utilize this
method decreases quickly with increasing redshift.}. The results
presented here should be cause for optimism that such progress can indeed be
made with continued advances in X-ray instrumentation, \eg \astroh\
(\citealt{ASTROH_TMP}) and \athena\ (\citealt{AthenaPlus, AthenaPlusSpin}), and
large observing programs, particularly if complemented by sensitive hard X-ray 
coverage similar to that provided by \nustar.

\section{Conclusions}
\label{sec_conc}

We present the first results from the full joint observing campaign undertaken
by \nustar\ and \xmm\ on the well known Seyfert galaxy \ngc. The four
coordinated observations reveal an extreme level of spectral variability, which
primarily appears to be due to variable line-of-sight absorption. However, while
changes in absorption have been observed in \ngc\ before (\eg \citealt{Risaliti05a,
Connolly14}), our observations display relatively unobscured states with high S/N
for the first time. Despite the diverse range of absorption states displayed, each
of the observations displays the same hallmarks of relativistic reflection from the
inner accretion disk: a relativistically broadened iron line and a strong Compton
reflection hump. These features therefore cannot be associated with line-of-sight
absorption. Indeed, with a simple phenomenological analysis, we find that the
strength of the relativistic iron line and the Compton hump relative to the intrinsic
continuum are well correlated, despite treating each of these features
independently, as expected if they are two manifestations of the same broadband
reflection spectrum.

Following this simple analysis, we perform time-resolved spectroscopy with
physically self-consistent disk reflection models, allowing for variable, partially
covering absorption, in order to constrain the inner disk parameters. We treat
each of the four observations independently in order to test the consistency
of the results obtained for the key physical parameters, \ie the black hole spin
and disk inclination, which should not vary on observable timescales. Excellent
consistency between these results obtained for these parameters from each of
the four observations is found. Despite the recent controversial claim by
\cite{LMiller13} that AGN spin cannot be measured, our results further
demonstrate that it is possible to measure the spin of the black holes powering
active galaxies, and we find that the central black hole in \ngc\ is indeed rapidly
rotating.

\section*{ACKNOWLEDGEMENTS}

The authors would like to thank the referee for providing useful feedback, which
helped improve the manuscript. This research has made use of data obtained
with the \nustar\ mission, a project led by the California Institute of Technology
(Caltech), managed by the Jet Propulsion Laboratory (JPL)  and funded by NASA,
and \xmm, an ESA science mission with instruments and contributions directly
funded by ESA Member States and NASA. We thank both the \xmm\ and the
\nustar\ Operations, Software and Calibration teams for support with the
execution and analysis of these coordinated observations. This research was
supported under NASA grant No. NNG08FD60C and has made use of the \nustar\
Data Analysis Software (\nustardas) jointly developed by the ASI Science Data
Center (ASDC, Italy) and Caltech (USA). PA acknowledges financial support from
Conicyt ACT 1101, and PG acknowledges support STFC (grant reference
ST/J00369711).

{\it Facilites:} \facility{NuSTAR}, \facility{XMM}

\bibliographystyle{/Users/dwalton/papers/mnras}

\bibliography{/Users/dwalton/papers/references}

\begin{thebibliography}{100}
\expandafter\ifx\csname natexlab\endcsname\relax\def\natexlab#1{#1}\fi

\bibitem[{Arnaud}(1996)]{XSPEC}
{Arnaud} K.~A., 1996, in { Astronomical Data Analysis Software and Systems
  V\/}, edited by {G.~H.~Jacoby \& J.~Barnes}, vol. 101 of { Astronomical
  Society of the Pacific Conference Series\/}, ~17

\bibitem[{Ballantyne} et~al.(2011){Ballantyne}, {McDuffie} \&
  {Rusin}]{Ballantyne11}
{Ballantyne} D.~R., {McDuffie} J.~R., {Rusin} J.~S., 2011, \apj, 734, 112

\bibitem[{Ballantyne} et~al.(2001){Ballantyne}, {Ross} \&
  {Fabian}]{Ballantyne01}
{Ballantyne} D.~R., {Ross} R.~R., {Fabian} A.~C., 2001, \mnras, 327, 10

\bibitem[{Ballantyne} et~al.(2004){Ballantyne}, {Turner} \&
  {Blaes}]{Ballantyne04}
{Ballantyne} D.~R., {Turner} N.~J., {Blaes} O.~M., 2004, \apj, 603, 436

\bibitem[{Berti} \& {Volonteri}(2008)]{Berti08}
{Berti} E., {Volonteri} M., 2008, \apj, 684, 822

\bibitem[{Bianchi} et~al.(2009){Bianchi}, {Guainazzi}, {Matt}, {Fonseca
  Bonilla} \& {Ponti}]{Bianchi09}
{Bianchi} S., {Guainazzi} M., {Matt} G., {Fonseca Bonilla} N., {Ponti} G.,
  2009, \aap, 495, 421

\bibitem[{Blandford} \& {Znajek}(1977)]{BZ77}
{Blandford} R.~D., {Znajek} R.~L., 1977, \mnras, 179, 433

\bibitem[{Blustin} et~al.(2005){Blustin}, {Page}, {Fuerst},
  {Branduardi-Raymont} \& {Ashton}]{Blustin05}
{Blustin} A.~J., {Page} M.~J., {Fuerst} S.~V., {Branduardi-Raymont} G.,
  {Ashton} C.~E., 2005, \aap, 431, 111

\bibitem[{Brenneman}(2013)]{Brenneman13book}
{Brenneman} L., 2013, {Measuring the Angular Momentum of Supermassive Black
  Holes}

\bibitem[{Brenneman} \& {Reynolds}(2006)]{kerrconv}
{Brenneman} L.~W., {Reynolds} C.~S., 2006, \apj, 652, 1028

\bibitem[{Brenneman} et~al.(2011){Brenneman}, {Reynolds}, {Nowak}
  et~al.]{Brenneman11}
{Brenneman} L.~W., {Reynolds} C.~S., {Nowak} M.~A., et~al., 2011, \apj, 736,
  103

\bibitem[{Brenneman} et~al.(2013){Brenneman}, {Risaliti}, {Elvis} \&
  {Nardini}]{Brenneman13}
{Brenneman} L.~W., {Risaliti} G., {Elvis} M., {Nardini} E., 2013, \mnras, 429,
  2662

\bibitem[{Cackett} et~al.(2013){Cackett}, {Fabian}, {Zogbhi}, {Kara},
  {Reynolds} \& {Uttley}]{Cackett13}
{Cackett} E.~M., {Fabian} A.~C., {Zogbhi} A., {Kara} E., {Reynolds} C.,
  {Uttley} P., 2013, \apjl, 764, L9

\bibitem[{Connolly} et~al.(2014){Connolly}, {McHardy} \& {Dwelly}]{Connolly14}
{Connolly} S.~D., {McHardy} I.~M., {Dwelly} T., 2014, ArXiv 1403.4253

\bibitem[{Cowperthwaite} \& {Reynolds}(2012)]{Cowperthwaite12}
{Cowperthwaite} P.~S., {Reynolds} C.~S., 2012, \apjl, 752, L21

\bibitem[{Dadina}(2007)]{Dadina07}
{Dadina} M., 2007, \aap, 461, 1209

\bibitem[{Dai} et~al.(2010){Dai}, {Kochanek}, {Chartas} et~al.]{Dai10}
{Dai} X., {Kochanek} C.~S., {Chartas} G., et~al., 2010, \apj, 709, 278

\bibitem[{Dauser} et~al.(2013){Dauser}, {Garcia}, {Wilms} et~al.]{Dauser13}
{Dauser} T., {Garcia} J., {Wilms} J., et~al., 2013, \mnras, 430, 1694

\bibitem[{Dauser} et~al.(2010){Dauser}, {Wilms}, {Reynolds} \&
  {Brenneman}]{relconv}
{Dauser} T., {Wilms} J., {Reynolds} C.~S., {Brenneman} L.~W., 2010, \mnras,
  409, 1534

\bibitem[{de La Calle P{\'e}rez} et~al.(2010){de La Calle P{\'e}rez},
  {Longinotti}, {Guainazzi} et~al.]{dLCPerez10}
{de La Calle P{\'e}rez} I., {Longinotti} A.~L., {Guainazzi} M., et~al., 2010,
  \aap, 524, A50

\bibitem[{De Marco} et~al.(2013){De Marco}, {Ponti}, {Cappi} et~al.]{deMarco13}
{De Marco} B., {Ponti} G., {Cappi} M., et~al., 2013, \mnras, 431, 2441

\bibitem[{Done} et~al.(2013){Done}, {Jin}, {Middleton} \& {Ward}]{Done13}
{Done} C., {Jin} C., {Middleton} M., {Ward} M., 2013, \mnras, 434, 1955

\bibitem[{Dovciak} et~al.(2013){Dovciak}, {Matt}, {Bianchi}
  et~al.]{AthenaPlusSpin}
{Dovciak} M., {Matt} G., {Bianchi} S., et~al., 2013, ArXiv 1306.2331

\bibitem[{Dubois} et~al.(2013){Dubois}, {Volonteri} \& {Silk}]{Dubois13}
{Dubois} Y., {Volonteri} M., {Silk} J., 2013, ArXiv 1304.4583

\bibitem[{Emmanoulopoulos} et~al.(2011){Emmanoulopoulos}, {McHardy} \&
  {Papadakis}]{Emman11b}
{Emmanoulopoulos} D., {McHardy} I.~M., {Papadakis} I.~E., 2011, \mnras, 416,
  L94

\bibitem[{Fabian} et~al.(2013){Fabian}, {Kara}, {Walton} et~al.]{Fabian13iras}
{Fabian} A.~C., {Kara} E., {Walton} D.~J., et~al., 2013, \mnras, 429, 2917

\bibitem[{Fabian} et~al.(1989){Fabian}, {Rees}, {Stella} \& {White}]{Fabian89}
{Fabian} A.~C., {Rees} M.~J., {Stella} L., {White} N.~E., 1989, \mnras, 238,
  729

\bibitem[{Fabian} et~al.(2009){Fabian}, {Zoghbi}, {Ross} et~al.]{FabZog09}
{Fabian} A.~C., {Zoghbi} A., {Ross} R.~R., et~al., 2009, \nat, 459, 540

\bibitem[{Gallo} et~al.(2013){Gallo}, {Fabian}, {Grupe} et~al.]{Gallo13}
{Gallo} L.~C., {Fabian} A.~C., {Grupe} D., et~al., 2013, \mnras, 428, 1191

\bibitem[{Garc{\'{\i}}a} \& {Kallman}(2010)]{xillver}
{Garc{\'{\i}}a} J., {Kallman} T.~R., 2010, \apj, 718, 695

\bibitem[{George} \& {Fabian}(1991)]{George91}
{George} I.~M., {Fabian} A.~C., 1991, \mnras, 249, 352

\bibitem[{Harrison} et~al.(2013){Harrison}, {Craig}, {Christensen}
  et~al.]{NUSTAR}
{Harrison} F.~A., {Craig} W.~W., {Christensen} F.~E., et~al., 2013, \apj, 770,
  103

\bibitem[{Jansen} et~al.(2001){Jansen}, {Lumb}, {Altieri} et~al.]{XMM}
{Jansen} F., {Lumb} D., {Altieri} B., et~al., 2001, \aap, 365, L1

\bibitem[{Kallman} \& {Bautista}(2001)]{xstar}
{Kallman} T., {Bautista} M., 2001, \apjs, 133, 221

\bibitem[{Kara} et~al.(2013){Kara}, {Fabian}, {Cackett}, {Uttley}, {Wilkins} \&
  {Zoghbi}]{Kara13feK}
{Kara} E., {Fabian} A.~C., {Cackett} E.~M., {Uttley} P., {Wilkins} D.~R.,
  {Zoghbi} A., 2013, \mnras, 434, 1129

\bibitem[{Kaspi} et~al.(2005){Kaspi}, {Maoz}, {Netzer}, {Peterson},
  {Vestergaard} \& {Jannuzi}]{Kaspi05}
{Kaspi} S., {Maoz} D., {Netzer} H., {Peterson} B.~M., {Vestergaard} M.,
  {Jannuzi} B.~T., 2005, \apj, 629, 61

\bibitem[{King} et~al.(2013){King}, {Miller}, {G{\"u}ltekin} et~al.]{King13jet}
{King} A.~L., {Miller} J.~M., {G{\"u}ltekin} K., et~al., 2013, \apj, 771, 84

\bibitem[{King} et~al.(2014){King}, {Walton}, {Miller} et~al.]{King14}
{King} A.~L., {Walton} D.~J., {Miller} J.~M., et~al., 2014, \apjl, 784, L2

\bibitem[{Kotov} et~al.(2001){Kotov}, {Churazov} \& {Gilfanov}]{Kotov01}
{Kotov} O., {Churazov} E., {Gilfanov} M., 2001, \mnras, 327, 799

\bibitem[{Laor}(1991)]{kdblur}
{Laor} A., 1991, \apj, 376, 90

\bibitem[{Legg} et~al.(2012){Legg}, {Miller}, {Turner}, {Giustini}, {Reeves} \&
  {Kraemer}]{Legg12}
{Legg} E., {Miller} L., {Turner} T.~J., {Giustini} M., {Reeves} J.~N.,
  {Kraemer} S.~B., 2012, \apj, 760, 73

\bibitem[{Lohfink} et~al.(2013){Lohfink}, {Reynolds}, {Jorstad}
  et~al.]{Lohfink13}
{Lohfink} A.~M., {Reynolds} C.~S., {Jorstad} S.~G., et~al., 2013, \apj, 772, 83

\bibitem[{Lohfink} et~al.(2012){Lohfink}, {Reynolds}, {Miller}
  et~al.]{Lohfink12}
{Lohfink} A.~M., {Reynolds} C.~S., {Miller} J.~M., et~al., 2012, \apj, 758, 67

\bibitem[{Magdziarz} \& {Zdziarski}(1995)]{pexrav}
{Magdziarz} P., {Zdziarski} A.~A., 1995, \mnras, 273, 837

\bibitem[{Maiolino} et~al.(2010){Maiolino}, {Risaliti}, {Salvati}
  et~al.]{Maiolino10}
{Maiolino} R., {Risaliti} G., {Salvati} M., et~al., 2010, \aap, 517, A47

\bibitem[{Marinucci} et~al.(2014{\natexlab{a}}){Marinucci}, {Matt}, {Kara}
  et~al.]{Marinucci14}
{Marinucci} A., {Matt} G., {Kara} E., et~al., 2014{\natexlab{a}}, \mnras, 440,
  2347

\bibitem[{Marinucci} et~al.(2014{\natexlab{b}}){Marinucci}, {Matt}, {Miniutti}
  et~al.]{Marinucci14mcg6}
{Marinucci} A., {Matt} G., {Miniutti} G., et~al., 2014{\natexlab{b}}, ArXiv
  1404.3561

\bibitem[{Miller}(2007)]{Miller07rev}
{Miller} J.~M., 2007, \araa, 45, 441

\bibitem[{Miller} et~al.(2011){Miller}, {Miller} \& {Reynolds}]{Miller11}
{Miller} J.~M., {Miller} M.~C., {Reynolds} C.~S., 2011, \apjl, 731, L5

\bibitem[{Miller} et~al.(2013){Miller}, {Parker}, {Fuerst} et~al.]{Miller13grs}
{Miller} J.~M., {Parker} M.~L., {Fuerst} F., et~al., 2013, \apjl, 775, L45

\bibitem[{Miller} et~al.(2012){Miller}, {Pooley}, {Fabian} et~al.]{Miller12}
{Miller} J.~M., {Pooley} G.~G., {Fabian} A.~C., et~al., 2012, \apj, 757, 11

\bibitem[{Miller} \& {Turner}(2013)]{LMiller13}
{Miller} L., {Turner} T.~J., 2013, \apjl, 773, L5

\bibitem[{Miller} et~al.(2008){Miller}, {Turner} \& {Reeves}]{Miller08}
{Miller} L., {Turner} T.~J., {Reeves} J.~N., 2008, \aap, 483, 437

\bibitem[{Miller} et~al.(2009){Miller}, {Turner} \& {Reeves}]{LMiller09}
{Miller} L., {Turner} T.~J., {Reeves} J.~N., 2009, \mnras, 399, L69

\bibitem[{Miller} et~al.(2010){Miller}, {Turner}, {Reeves} \&
  {Braito}]{LMiller10b}
{Miller} L., {Turner} T.~J., {Reeves} J.~N., {Braito} V., 2010, \mnras, 408,
  1928

\bibitem[{Miniutti} \& {Fabian}(2004)]{lightbending}
{Miniutti} G., {Fabian} A.~C., 2004, \mnras, 349, 1435

\bibitem[{Miniutti} et~al.(2009){Miniutti}, {Panessa}, {de Rosa}
  et~al.]{Miniutti09sw2127}
{Miniutti} G., {Panessa} F., {de Rosa} A., et~al., 2009, \mnras, 398, 255

\bibitem[{Nandra} et~al.(2013){Nandra}, {Barret}, {Barcons} et~al.]{AthenaPlus}
{Nandra} K., {Barret} D., {Barcons} X., et~al., 2013, ArXiv e-prints

\bibitem[{Nandra} et~al.(2007){Nandra}, {O'Neill}, {George} \&
  {Reeves}]{Nandra07}
{Nandra} K., {O'Neill} P.~M., {George} I.~M., {Reeves} J.~N., 2007, \mnras,
  382, 194

\bibitem[{Nandra} \& {Pounds}(1994)]{Nandra94}
{Nandra} K., {Pounds} K.~A., 1994, \mnras, 268, 405

\bibitem[{Nardini} et~al.(2011){Nardini}, {Fabian}, {Reis} \&
  {Walton}]{Nardini11}
{Nardini} E., {Fabian} A.~C., {Reis} R.~C., {Walton} D.~J., 2011, \mnras, 410,
  1251

\bibitem[{Nardini} et~al.(2012){Nardini}, {Fabian} \& {Walton}]{Nardini12}
{Nardini} E., {Fabian} A.~C., {Walton} D.~J., 2012, \mnras, 423, 3299

\bibitem[{Nayakshin} et~al.(2000){Nayakshin}, {Kazanas} \&
  {Kallman}]{Nayakshin00}
{Nayakshin} S., {Kazanas} D., {Kallman} T.~R., 2000, \apj, 537, 833

\bibitem[{Parker} et~al.(2014{\natexlab{a}}){Parker}, {Marinucci}, {Brenneman}
  et~al.]{Parker14mcg}
{Parker} M.~L., {Marinucci} A., {Brenneman} L., et~al., 2014{\natexlab{a}},
  \mnras, 437, 721

\bibitem[{Parker} et~al.(2014{\natexlab{b}}){Parker}, {Walton}, {Fabian} \&
  {Risaliti}]{Parker14}
{Parker} M.~L., {Walton} D.~J., {Fabian} A.~C., {Risaliti} G.,
  2014{\natexlab{b}}, ArXiv e-prints

\bibitem[{Perola} et~al.(2002){Perola}, {Matt}, {Cappi} et~al.]{Perola02}
{Perola} G.~C., {Matt} G., {Cappi} M., et~al., 2002, \aap, 389, 802

\bibitem[{Reis} et~al.(2012{\natexlab{a}}){Reis}, {Fabian}, {Reynolds}
  et~al.]{Reis3783}
{Reis} R.~C., {Fabian} A.~C., {Reynolds} C.~S., et~al., 2012{\natexlab{a}},
  \apj, 745, 93

\bibitem[{Reis} et~al.(2012{\natexlab{b}}){Reis}, {Miller}, {Reynolds},
  {Fabian} \& {Walton}]{Reis1836}
{Reis} R.~C., {Miller} J.~M., {Reynolds} M.~T., {Fabian} A.~C., {Walton} D.~J.,
  2012{\natexlab{b}}, \apj, 751, 34

\bibitem[{Reis} et~al.(2014){Reis}, {Reynolds}, {Miller} \&
  {Walton}]{Reis14nat}
{Reis} R.~C., {Reynolds} M.~T., {Miller} J.~M., {Walton} D.~J., 2014, \nat,
  507, 207

\bibitem[{Reynolds}(2013)]{Reynolds13rev}
{Reynolds} C.~S., 2013, \ssr

\bibitem[{Reynolds} \& {Fabian}(2008)]{Reynolds08}
{Reynolds} C.~S., {Fabian} A.~C., 2008, \apj, 675, 1048

\bibitem[{Risaliti} et~al.(2005{\natexlab{a}}){Risaliti}, {Bianchi}, {Matt}
  et~al.]{Risaliti05b}
{Risaliti} G., {Bianchi} S., {Matt} G., et~al., 2005{\natexlab{a}}, \apjl, 630,
  L129

\bibitem[{Risaliti} et~al.(2005{\natexlab{b}}){Risaliti}, {Elvis}, {Fabbiano},
  {Baldi} \& {Zezas}]{Risaliti05a}
{Risaliti} G., {Elvis} M., {Fabbiano} G., {Baldi} A., {Zezas} A.,
  2005{\natexlab{b}}, \apjl, 623, L93

\bibitem[{Risaliti} et~al.(2002){Risaliti}, {Elvis} \& {Nicastro}]{Risaliti02}
{Risaliti} G., {Elvis} M., {Nicastro} F., 2002, \apj, 571, 234

\bibitem[{Risaliti} et~al.(2013){Risaliti}, {Harrison}, {Madsen}
  et~al.]{Risaliti13nat}
{Risaliti} G., {Harrison} F.~A., {Madsen} K.~K., et~al., 2013, \nat, 494, 449

\bibitem[{Risaliti} et~al.(2009){Risaliti}, {Miniutti}, {Elvis}
  et~al.]{Risaliti09a}
{Risaliti} G., {Miniutti} G., {Elvis} M., et~al., 2009, \apj, 696, 160

\bibitem[{Rivers} et~al.(2013){Rivers}, {Markowitz} \& {Rothschild}]{Rivers13}
{Rivers} E., {Markowitz} A., {Rothschild} R., 2013, \apj, 772, 114

\bibitem[{Ross} \& {Fabian}(2005)]{reflion}
{Ross} R.~R., {Fabian} A.~C., 2005, \mnras, 358, 211

\bibitem[{Russell} et~al.(2013){Russell}, {Gallo} \& {Fender}]{Russell13}
{Russell} D.~M., {Gallo} E., {Fender} R.~P., 2013, \mnras, 431, 405

\bibitem[{Schulz} et~al.(1999){Schulz}, {Komossa}, {Schmitz} \&
  {M{\"u}cke}]{Schulz99}
{Schulz} H., {Komossa} S., {Schmitz} C., {M{\"u}cke} A., 1999, \aap, 346, 764

\bibitem[{Sim} et~al.(2010){Sim}, {Proga}, {Miller}, {Long} \& {Turner}]{Sim10}
{Sim} S.~A., {Proga} D., {Miller} L., {Long} K.~S., {Turner} T.~J., 2010,
  \mnras, 408, 1396

\bibitem[{Steiner} et~al.(2013){Steiner}, {McClintock} \& {Narayan}]{Steiner13}
{Steiner} J.~F., {McClintock} J.~E., {Narayan} R., 2013, \apj, 762, 104

\bibitem[{Steiner} et~al.(2012){Steiner}, {Reis}, {Fabian}
  et~al.]{Steiner12lmc}
{Steiner} J.~F., {Reis} R.~C., {Fabian} A.~C., et~al., 2012, \mnras, 427, 2552

\bibitem[{Str{\"u}der} et~al.(2001){Str{\"u}der}, {Briel}, {Dennerl}
  et~al.]{XMM_PN}
{Str{\"u}der} L., {Briel} U., {Dennerl} K., et~al., 2001, \aap, 365, L18

\bibitem[{Takahashi} et~al.(2012){Takahashi}, {Mitsuda}, {Kelley}
  et~al.]{ASTROH_TMP}
{Takahashi} T., {Mitsuda} K., {Kelley} R., et~al., 2012, in { Society of
  Photo-Optical Instrumentation Engineers (SPIE) Conference Series\/}, vol.
  8443 of { Society of Photo-Optical Instrumentation Engineers (SPIE)
  Conference Series\/}

\bibitem[{Tatum} et~al.(2013){Tatum}, {Turner}, {Miller} \& {Reeves}]{Tatum13}
{Tatum} M.~M., {Turner} T.~J., {Miller} L., {Reeves} J.~N., 2013, \apj, 762, 80

\bibitem[{Tomsick} et~al.(2014){Tomsick}, {Nowak}, {Parker} et~al.]{Tomsick14}
{Tomsick} J.~A., {Nowak} M.~A., {Parker} M., et~al., 2014, \apj, 780, 78

\bibitem[{Turner} et~al.(2001){Turner}, {Abbey}, {Arnaud} et~al.]{XMM_MOS}
{Turner} M.~J.~L., {Abbey} A., {Arnaud} M., et~al., 2001, \aap, 365, L27

\bibitem[{Walton} et~al.(2013{\natexlab{a}}){Walton}, {Fuerst}, {Harrison}
  et~al.]{Walton13culx}
{Walton} D.~J., {Fuerst} F., {Harrison} F., et~al., 2013{\natexlab{a}}, \apj,
  779, 148

\bibitem[{Walton} et~al.(2014){Walton}, {Harrison}, {Grefenstette}
  et~al.]{Walton14hoIX}
{Walton} D.~J., {Harrison} F.~A., {Grefenstette} B.~W., et~al., 2014, ArXiv
  1402.2992

\bibitem[{Walton} et~al.(2013{\natexlab{b}}){Walton}, {Nardini}, {Fabian},
  {Gallo} \& {Reis}]{Walton13spin}
{Walton} D.~J., {Nardini} E., {Fabian} A.~C., {Gallo} L.~C., {Reis} R.~C.,
  2013{\natexlab{b}}, \mnras, 428, 2901

\bibitem[{Walton} et~al.(2012){Walton}, {Reis}, {Cackett}, {Fabian} \&
  {Miller}]{Walton12xrbAGN}
{Walton} D.~J., {Reis} R.~C., {Cackett} E.~M., {Fabian} A.~C., {Miller} J.~M.,
  2012, \mnras, 422, 2510

\bibitem[{Walton} et~al.(2010){Walton}, {Reis} \& {Fabian}]{Walton10Hex}
{Walton} D.~J., {Reis} R.~C., {Fabian} A.~C., 2010, \mnras, 408, 601

\bibitem[{Walton} et~al.(2013{\natexlab{c}}){Walton}, {Zoghbi}, {Cackett}
  et~al.]{Walton13lag}
{Walton} D.~J., {Zoghbi} A., {Cackett} E.~M., et~al., 2013{\natexlab{c}},
  \apjl, 777, L23

\bibitem[{Wang} et~al.(2009){Wang}, {Fabbiano}, {Elvis} et~al.]{Wang09}
{Wang} J., {Fabbiano} G., {Elvis} M., et~al., 2009, \apj, 694, 718

\bibitem[{Wilkins} \& {Fabian}(2012)]{Wilkins12}
{Wilkins} D.~R., {Fabian} A.~C., 2012, \mnras, 424, 1284

\bibitem[{Zhang} et~al.(1997){Zhang}, {Cui} \& {Chen}]{Zhang97}
{Zhang} S.~N., {Cui} W., {Chen} W., 1997, \apjl, 482, L155

\bibitem[{Zoghbi} et~al.(2012){Zoghbi}, {Fabian}, {Reynolds} \&
  {Cackett}]{Zoghbi12}
{Zoghbi} A., {Fabian} A.~C., {Reynolds} C.~S., {Cackett} E.~M., 2012, \mnras,
  422, 129

\bibitem[{Zoghbi} et~al.(2010){Zoghbi}, {Fabian}, {Uttley} et~al.]{Zoghbi10}
{Zoghbi} A., {Fabian} A.~C., {Uttley} P., et~al., 2010, \mnras, 401, 2419

\bibitem[{Zoghbi} et~al.(2013){Zoghbi}, {Reynolds}, {Cackett}, {Miniutti},
  {Kara} \& {Fabian}]{Zoghbi13a}
{Zoghbi} A., {Reynolds} C., {Cackett} E.~M., {Miniutti} G., {Kara} E., {Fabian}
  A.~C., 2013, \apj, 767, 121

\end{thebibliography}

\label{lastpage}

\end{document}